\documentclass[aps,nobalancelastpage, prb,twocolumn,footinbib, superscriptaddress]{revtex4-2}

\usepackage{graphicx}
\usepackage{amsmath}
\usepackage{amssymb}
\usepackage{bbm}
\usepackage{float}
\usepackage[colorlinks=True,linkcolor=blue,citecolor=blue,urlcolor=blue]{hyperref}
\usepackage{amsfonts}
\usepackage{braket}
\usepackage[dvipsnames]{xcolor}
\begin{document}

\title{Probability transport on the Fock space of a disordered quantum spin chain}


\newcommand{\comment}[1]{}
\newcommand{\pd}{\phantom\dagger}

\newcommand{\dt}{\tilde \delta}
\newcommand{\ua}{\uparrow}
\newcommand{\da}{\downarrow}
\newcommand{\w}{\omega}

\newcommand{\wtil}{\tilde{\w}}
\newcommand{\wtilp}{\tilde{\w}^{\prime}}
\newcommand{\ecal}{{\cal{E}}}

\newcommand{\n}[2][I]{n^{\scriptscriptstyle(#1)}_{#2}}
\newcommand{\Ntot}[2][I]{N^{\scriptscriptstyle(#1)}_{#2}}
\newcommand{\Sump}{\sum^{\prime}}

\newcommand{\nh}{N_{\mathcal{H}}}
\newcommand{\nd}{\mathcal{D}}

\newcommand{\mblo}{\mathrm{MBL}_{0}}

\newcommand{\PIJn}[2][n]{P^{\scriptscriptstyle(#1)}_{#2}}
\newcommand{\rn}[2][n]{r^{\scriptscriptstyle(#1)}_{#2}}
\newcommand{\barrn}[2][n]{\overline{r}^{\scriptscriptstyle(#1)}_{#2}}
\newcommand{\Cn}[2][n]{\mathcal{C}^{\scriptscriptstyle[#1]}_{#2}}
\newcommand{\barCn}[2][n]{\overline{\mathcal{C}}^{\scriptscriptstyle[#1]}}
\newcommand{\Clln}[2][n]{\mathcal{C}_{\ell\ell}^{\scriptscriptstyle[#1]}}
\newcommand{\CllI}[2][I]{\mathcal{C}_{\ell\ell}^{\scriptscriptstyle[#1]}}
\newcommand{\barClln}[2][n]{\overline{\mathcal{C}}_{\ell\ell}^{\scriptscriptstyle[#1]}}
\newcommand{\AJI}[2][I]{\mathcal{A}^{\scriptscriptstyle(#1)}_{#2}}
\newcommand{\BJI}[2][I]{\mathcal{B}^{\scriptscriptstyle(#1)}_{#2}}

\newcommand{\wc}{W_{\mathrm{c}}}

\newcommand{\prel}{P_{\mathrm{rel}}}
\newcommand{\prelc}{P_{\mathrm{rel,c}}}

\newcommand\eq[1]{\begin{align}#1\end{align}}
\newcommand{\C}{\mathcal{C}}
\newcommand{\tCn}{\tilde{\mathcal{C}}^{[n]}}
\newcommand{\M}{\overline{\mathcal{M}}}
\newcommand{\Ss}{\mathcal{S}}
\newcommand{\Mi}{\mathcal{M}_{i}}
\newcommand{\Ms}{\mathcal{M}_\mathrm{S}}
\newcommand{\Ci}{\mathcal{C}^{[I]}}
\newcommand{\inter}{\chi_\mathrm{inter}}
\newcommand{\intra}{\chi_\mathrm{intra}}
\newcommand{\ai}{|A_{I}|^2}
\newcommand{\ipr}{\overline{\mathcal{L}_2}}
\newcommand{\ltwo}{\mathcal{L}_2}
\newcommand{\mbln}{\mathrm{MBL}_0}
\newcommand{\Fr}{\mathcal{F}}
\newcommand{\Fd}{\mathcal{F}_d}
\newcommand{\Fs}{\mathcal{F}_s}
\newcommand{\R}{\mathcal{R}_2}
\newcommand{\D}{\mathcal{D}}
\newcommand{\emax}{\tilde{E}_{\mathrm{max}}}
\newcommand{\ttil}{\tilde{t}}
\newcommand{\ttilT}{\tilde{t}_{\mathrm{Th}}}
\newcommand{\st}{S_{\mathrm{tot}}^{z}}
\newcommand{\tm}{t_{\mathrm{m}}}
\newcommand{\tH}{t_{\mathrm{H}}}
\newcommand{\dis}{{\mathrm{d}}}
\newcommand{\rel}{{\mathrm{rel}}}

\author{Isabel Creed}
\email{isabel.creed@chem.ox.ac.uk}
\affiliation{Physical and Theoretical Chemistry, Oxford University, 
South Parks Road, Oxford OX1 3QZ, United Kingdom}
\author{David E. Logan}
\email{david.logan@chem.ox.ac.uk}
\affiliation{Physical and Theoretical Chemistry, Oxford University,
South Parks Road, Oxford OX1 3QZ, United Kingdom}
\affiliation{Department of Physics, Indian Institute of Science, Bengaluru 560012, India}
\author{Sthitadhi Roy}
\email{sthitadhi.roy@icts.res.in}
\affiliation{International Centre for Theoretical Sciences, Tata Institute of Fundamental Research, Bengaluru, 560089, India  }

\date{\today}

\begin{abstract}
Within the broad theme of understanding the dynamics of disordered  quantum many-body systems, one of the simplest questions one can ask is: given an initial state, how does it evolve in time on the associated Fock-space graph, in terms of the distribution of probabilities thereon? A detailed quantitative description of the temporal evolution of out-of-equilibrium disordered quantum states and probability transport on the Fock space, is our central aim here. We investigate  it in the context of a disordered quantum spin chain which hosts a disorder-driven many-body localisation transition. Real-time dynamics/probability transport is shown to exhibit a rich phenomenology, which is markedly different between the ergodic and many-body localised phases. The dynamics is for example found to be strongly inhomogeneous at intermediate times   in both phases, but while it gives way to homogeneity at long times in the ergodic phase, the dynamics remain inhomogeneous and multifractal in nature for arbitrarily long times in the localised phase. Similarly, we show that an appropriately defined dynamical lengthscale on the Fock-space graph is directly related to the local spin-autocorrelation, and as such sheds light on the (anomalous) decay of the autocorrelation in the ergodic phase, and lack of it in the localised phase. 
\end{abstract}

\maketitle


\section{Introduction}
\label{section:intro}

The out-of-equilibrium dynamics of isolated quantum many-body systems can show a rich range of behaviour in the presence of disorder. One of the most striking examples is the driving of such a system from the default ergodic phase into a many-body localised (MBL) phase at sufficiently strong disorder~\cite{basko2006metal,gornyi2005interacting,oganesyan2007localisation,znidaric2008many,lev2015absence,nandkishore2015many,abanin2017recent,abanin2019colloquium}, via a dynamical phase transition~\cite{pal2010many,luitz2015many}. In contrast to the ergodic phase, the system in the MBL phase fails to thermalise under its own dynamics, and memory of the initial state survives locally for arbitrarily long times. Standard signatures of these include the absence of transport of conserved quantities, and autocorrelations of local observables  saturating to finite values at long times~\cite{znidaric2008many,serbyn2014quantum,luitz2016extended}, rather than vanishing. As such behaviour falls outside the paradigm of conventional statistical mechanics, the dynamics in the MBL phase is naturally of fundamental interest. At the same time, even within the ergodic phase but at disorder strengths preceding the MBL transition, the dynamics is anomalously slow. This is commonly manifest in subdiffusive transport of conserved quantities, and autocorrelations of local observables decaying in time with anomalous power-law exponents~\cite{agarwal2015anomalous,SantosPRB2015,luitz2016long,luitz2016anomalous,luitz2017ergodic,roy2018anomalous}. 
This behaviour attests to the fact that the out-of-equilibrium dynamics of disordered quantum systems across a range of disorder strengths straddling the MBL transition is an interesting question.

From a phenomenogical point of view, there has been substantial progress in understanding the dynamics, both in the MBL phase as well as in the anomalous ergodic regime. The absence of transport in the MBL phase can be explained via the presence of an extensive number of emergent local integrals of motion (or equivalently, local conserved charges), such that an effective model for the MBL phase involves interactions only between these entities~\cite{huse2014phenomenology,serbyn2013local,ros2015integrals}. More recently, resonances between configurations of these charges have been shown to further explain several features of the MBL phase~\cite{garratt2021local,MorningstarAvalanchesPRB2022,ghosh2022resonance,garratt2022resonant,crowley2022constructive,long2022prethermal}. In the anomalous ergodic regime, progress in understanding the slow dynamics has centred on phenomenological theories based on rare Griffiths regions, as well as anomalous spectral properties of local observables~\cite{agarwal2015anomalous,luitz2016anomalous,roy2018anomalous}. It is nevertheless desirable to have a theoretical framework, rooted in microscopics, for understanding both the slow dynamics preceding the MBL phase and the arrested dynamics in the MBL phase. Mapping the dynamics of the many-body system to that of probabilities on its Fock-space graph provides such a framework.

Indeed, understanding the physics of many-body localisation from the perspective of the
associated Fock space (FS) has emerged as a fruitful approach over the last few years ~\cite{altshuler1997quasiparticle,basko2006metal,MonthusGarel2010PRB,deluca2013ergodicity,serbyn2015criterion,pietracaprina2016forward,baldwin2016manybody,logan2019many,roy2018exact,roy2018percolation,roy2019self,mace2019multifractal,pietracaprina2021hilbert,detomasi2019dynamics,ghosh2019manybody,nag2019manybody,roy2020fock,biroli2020anomalous,tarzia2020manybody,detomasi2020rare,hopjan2020manybody,tikhonov2021eigenstate,roy2021fockspace,sutradhar2022scaling,roy2022hilbert}. This approach involves recasting the Hamiltonian of a disordered, interacting quantum 
system as a tight-binding Hamiltonian on the complex, correlated FS graph of the system~\cite{welsh2018simple}. The problem then becomes one of Anderson localisation (AL) of a fictitious particle on the FS graph, albeit a distinctly unconventional AL problem due to the strong correlations in effective disorder on the FS graph~\cite{roy2020fock}. This mapping opens a new window into the connections between the anatomy of the eigenstates on the FS graph, and their manifestations in terms of real-space properties. For instance, the spread of the eigenstates on the FS graph, and an associated emergent correlation length, has been shown to carry information about eigenstate expectation values of local observables~\cite{roy2021fockspace} as well as that of the $l$-bit localisation length~\cite{detomasi2020rare}. Higher-point correlations of eigenstate amplitudes encode their entanglement structure~\cite{roy2022hilbert}. However, most of these studies have focussed on eigenstate properties, and much less so in the context of dynamical, time-dependent properties.

Motivated by this, we investigate here the dynamics of an out-of-equilibrium quantum state on the FS graph. Arguably, the most fundamental question one can ask in this regard is: given an initial out-of-equilibrium state, how do the probability densities of the state on the FS graph evolve in time and spread out on the graph?  As will be shown, a detailed characterisation of this probability transport carries a plethora of information providing insights into the dynamics of disordered quantum many-body systems. This is the central goal of  the work. We begin with a brief overview of the paper.


\subsection*{Overview \label{sec:overview}}

As a concrete setting for our analysis, we consider a quantum Ising chain with disordered 
longitudinal fields and interactions, together with a constant transverse field (of strength 
$\Gamma$). A description of the model and its associated FS graph is given in Sec.~\ref{section:model}. Classical spin-configurations form a convenient set of basis states; they also form the nodes (or `sites') of the FS graph, with the transverse field generating links between them. The Hamming distance~\footnote{The Hamming distance between two classical spin configurations is simply the number of spins that are different between the two.} between two classical spin configurations endows the FS graph with a natural measure of distance. Initialising the state in a classical spin configuration corresponds to intialising it on a site on the FS graph. Consequently, the FS graph can be organised such that the given initial state sits at the apex and all sites at a fixed Hamming distance from the initial site are arranged row-wise (see Fig.~\ref{fig:fockspace}).

Although the FS graph for a chain of length $L$ is an $L$-dimensional hypercube, the above organisation of the graph gives rise to two natural `axes' along which the probability transport can be defined; we refer to them as \emph{longitudinal} and \emph{lateral} probability transport. The former quantifies how the probability flows down sites which are at increasing distances from the initial FS site. Lateral probability transport on the other hand  measures how the probability spreads across sites at the same Hamming distance from the initial site, 
i.e.\ on a given row. Sec.~\ref{section:probsintro} formalises these two notions of FS probability transport. We show in particular that a time-dependent lengthscale, 
$\overline{r}(t)$, which characterises the longitudinal spread of the wavefunction, is directly related to the real-space spin autocorrelation function. We also quantify the extent to which the time-evolving state is (de)localised on the graph, via $t$-dependent inverse participation ratios (IPR) and their corresponding fractal exponents. These IPRs can be defined over the entire FS graph, or can be defined row-wise (which corresponds to the lateral transport).

In Sec.~\ref{section:shortt} we analyse the short-time dynamics, which is independent of
whether the ultimate late-time behaviour of the system is ergodic or MBL in character. For 
$\Gamma t\ll 1$, the probability of finding the system in a given FS site/state at distance $r$ 
is shown to scale as $\sim (\Gamma t)^{2r}$.  An essential outcome  of this is an 
\emph{emergent multifractality} of the wavefunction over the full FS, with a fractal exponent growing $\propto t^2$, independent of disorder strength. By contrast, the row-resolved IPRs on these timescales do not show fractal statistics, indicating that the short-time wavefunction is spread homogeneously across any given row of the FS graph. A further, rather striking consequence of the analysis, is that $\overline{r}(t)$ becomes extensive in system size $L$ at any finite $\mathcal{O}(1)$ time. This is mandated by the spin autocorrelation being strictly $<1$ at any finite $\mathcal{O}(1)$ time, and can be understood via the extensive connectivity of the FS graph.

Section~\ref{section:verticaltransport}  is devoted to consideration of longitudinal probability transport, notably for long-times.  A central result here is that,  in the ergodic regime, the lengthscale $\overline{r}(t)$ grows subdiffusively, $\sim t^\alpha$ with $\alpha<1/2$, until 
it reaches its maximal value of $L/2$ (modulo the role of mobility edges and finite-size effects, as explained later). This is shown to imply that the spin-autocorrelation also decays as a power-law with the same exponent. In the MBL regime by contrast, $\overline{r}(t)$
saturates to an extensive but sub-maximal value, which in turn implies that the spin autocorrelation remains non-zero at arbitrarily long times. A further implication of these results is that the emergent fractality present at short to intermediate times gives way to fully delocalised states at long-times in the ergodic regime, whereas the fractality persists for arbitrarily long times in the MBL regime.

 In Sec.~\ref{section:lateraltransportmain} we turn to the analysis of lateral probability transport, via row-resolved IPRs. The picture which emerges is that, following the short-time homogeneity, at intermediate times -- and for any disorder strength -- the time-dependent probabilities on any row develop strong inhomogeneities, reflected in (multi)fractal scalings of the row-resolved IPRs. For sufficiently long times, however, this fractality gives way to complete homogeneity in the ergodic regime, while it persists in the MBL regime. Since the lateral transport in essence captures inhomogeneity in the evolution of probabilities on the FS graph, it is also natural to study $t$-dependent distributions of probabilities over sites on a given row. Consistent with the above picture, we find that the inhomogeneities are accompanied by heavy-tailed L\'evy distributions, whereas temporal regimes in which probability spreads homogeneously are characterised by narrow distributions.

 We summarise our results in Sec.~\ref{sec:summary} (see Fig.~\ref{fig:summary} for a visual 
summary), and close with concluding remarks and a future outlook.


\section{Model and Fock-space graph}
\label{section:model}

We consider a disordered transverse-field Ising (TFI) spin-1/2 chain, specified by
the Hamiltonian
\begin{equation}
\label{eq:ham}
\mathcal{H} ~=~ \sum_{\ell=1}^{L-1}J_\ell^{\pd}\hat{\sigma}^z_{\ell}\hat{\sigma}^z_{\ell+1}+ \sum_{\ell=1}^L[h_\ell^{\pd}\hat{\sigma}^z_\ell + \Gamma\hat{\sigma}^x_\ell]\,,
\end{equation}
where $h_\ell$ and $J_\ell$ are i.i.d.\ random variables, uniformly distributed with 
$h_\ell\in[-W,W]$ and $J_\ell\in[J-\delta J,J+\delta J]$. For numerical studies, we consider 
$J=1$, $\delta J=0.2$, and $\Gamma=1$.  With these parameters, and the range of system sizes accessible in practice to exact diagonalisation (ED), the critical disorder strength above which all eigenstates are MBL is estimated to be $\wc \simeq 3.8$~\cite{abanin2021distinguishing}.
Some recent works on standard disordered models~\cite{VidmarChaosPRE2020,suntajs2020ergodicity,sirker2020evidence,sels2021dynamical,sirker2021slow,MorningstarAvalanchesPRB2022,SelsPRB2022,long2022prethermal} have however suggested that a genuine MBL phase, stable in the  thermodynamic limit $L\to \infty$, can arise only for much larger values 
of $W$, and that the apparent localisation found for finite systems at $W > \wc$ is indicative of a  pre-thermal regime. Here we take the view that the MBL phenomenology clearly observed at 
$W>\wc \sim 4$ for ED-accessible system sizes persists in the thermodynamic limit, albeit for 
larger $W$ values.

\begin{figure}
\includegraphics[width=\columnwidth]{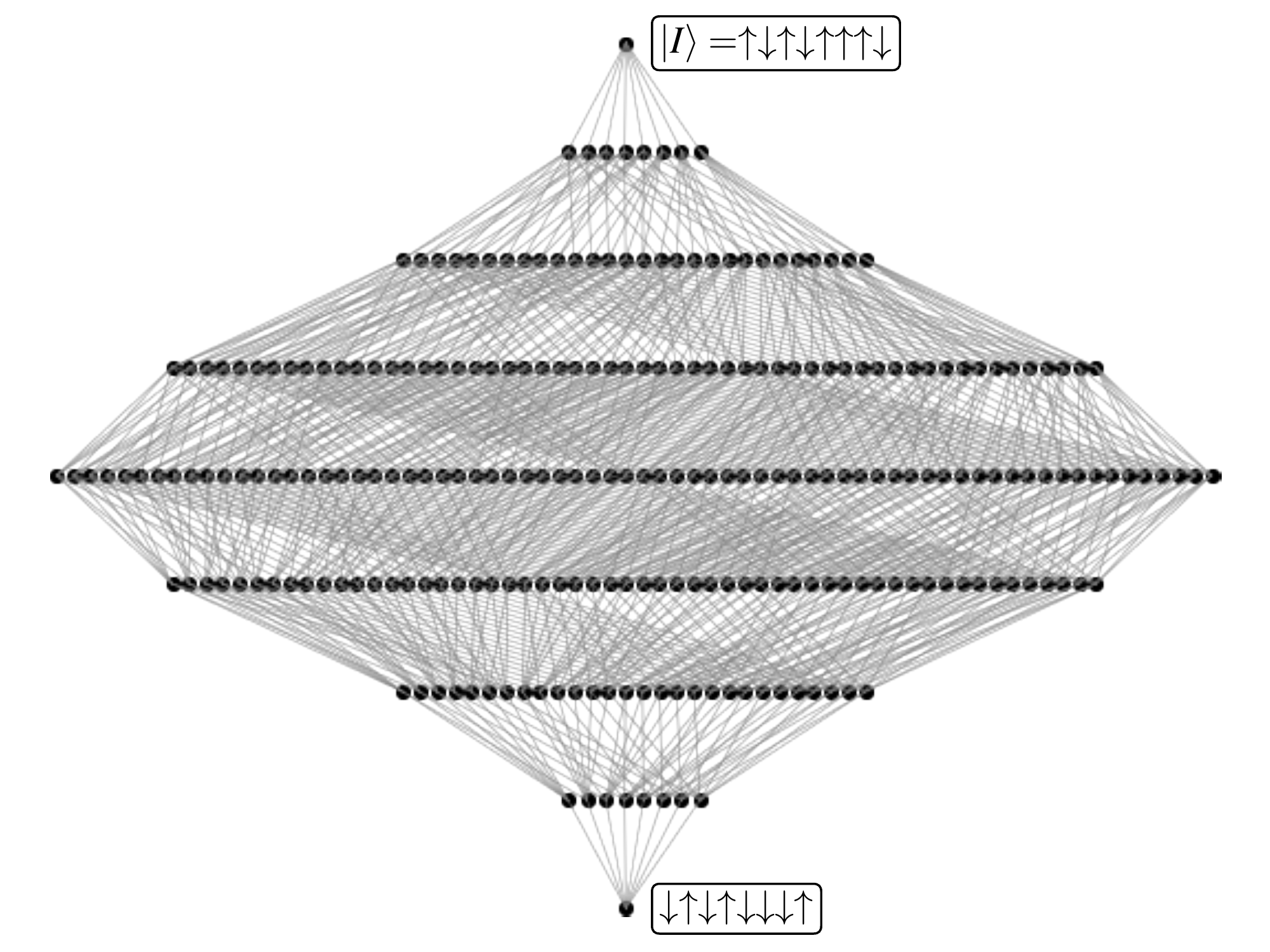}
\caption{Fock-space (FS) graph of the disordered TFI model Eq.~\ref{eq:ham} in the basis of 
$\sigma^z$-product states, illustrated for $L=8$. An arbitrary FS site, $I$,
is placed at the apex. The graph has $L+1$ rows, and the number of FS sites on row $r$ is 
$N_{r}=\binom{L}{r}$. Any FS site $J$ on row $r$ is a Hamming distance $r_{IJ}=r$ from $I$.
Links/hoppings can connect only FS sites on adjacent rows; with each FS site connected
to precisely $L$ others.
}
\label{fig:fockspace}
\end{figure}

Fock space (FS) provides a natural framework for studying many-body localisation~\cite{altshuler1997quasiparticle,MonthusGarel2010PRB,deluca2013ergodicity,serbyn2015criterion,pietracaprina2016forward,baldwin2016manybody,logan2019many,roy2018exact,roy2018percolation,roy2019self,mace2019multifractal,pietracaprina2021hilbert,detomasi2019dynamics,ghosh2019manybody,nag2019manybody,roy2020fock,biroli2020anomalous,tarzia2020manybody,detomasi2020rare,hopjan2020manybody,tikhonov2021eigenstate,roy2021fockspace,sutradhar2022scaling,roy2022hilbert}, 
in part because a generic many-body Hamiltonian maps exactly onto a tight-binding model on the associated FS graph (or `lattice'), of form
\begin{equation}
\label{eq:hamTBM}
\mathcal{H} ~=~ \sum_{J}\mathcal{E}_{J}^{\pd}|J\rangle\langle J| 
~+~{\sum_{J,K}}^{\prime} T_{JK}^{\pd}|J\rangle\langle K|
\end{equation}
(where $^{\prime}$ means $K\neq J$). The FS graph of the TFI model in the basis of 
$\sigma^z$-product states is an $L$-dimensional hypercube with $\nh=2^L$ vertices, or FS sites, as illustrated in Fig.~\ref{fig:fockspace}. A FS site, $J$, represents a many-body quantum state $|J\rangle$ of $L$ spins which is an eigenstate of each $\hat{\sigma}^z_\ell$-operator, $\hat{\sigma}^z_\ell\ket{J}=S_{\ell,J}\ket{J}$ where $S_{\ell,J}=\pm 1$.
It is thus an eigenstate of $\mathcal{H}_{0}=\sum_{\ell}[h_{\ell}\hat{\sigma}_{\ell}^{z} +J_{\ell}\hat{\sigma}_{\ell}^{z}\hat{\sigma}_{\ell+1}^{z}]$, i.e.\ $\mathcal{H}_{0}|J\rangle =\mathcal{E}_{J}|J\rangle$, with $\mathcal{E}_{J}$ the corresponding site-energy for the FS site
(with the $\{\mathcal{E}_{J}\}$ maximally correlated~\cite{roy2020fock}, and not i.i.d.).
Links, or hoppings, on the FS graph are generated by the term
$\mathcal{H}_{1}=\mathcal{H}-\mathcal{H}_{0}=\Gamma\sum_{\ell}\hat{\sigma}_{\ell}^{x}$. 
Each FS site is thus connected to precisely $L$ others, lying solely on adjacent rows of the
graph, and each of which corresponds to flipping a spin on a particular real-space site.
This generates the hopping contribution to Eq.\ \ref{eq:hamTBM},
in which all non-vanishing hopping matrix elements are simply $T_{JK}=\Gamma$.

As illustrated in Fig.~\ref{fig:fockspace} the graph consists of $L+1$ rows, $r=0-L$.
A single FS site, denoted by $I$ in Fig.~\ref{fig:fockspace} (and with arbitrary
spin orientations for the real-space sites) lies at the apex of the graph, $r=0$.
The number of FS sites on row $r$ of the graph is $N_{r}:=\binom{L}{r}$; with the final 
site, $r=L$, corresponding to the state $|\overline{I}\rangle$ in which all real-space spins on 
$|I\rangle$ have been flipped. As a measure of distance between two sites on the FS graph we use the Hamming distance, as mentioned in Sec.\ \ref{sec:overview}. For any pair of FS sites $J,I$ separated by a Hamming distance $r_{IJ}=r$, then by definition $r$ real-space sites $\ell$ have
$S_{\ell,J}=-S_{\ell,I}$ while $L-r$ sites have $S_{\ell,J}=+S_{\ell,I}$. Hence
\begin{equation}
	\label{eq:rIJ}
	L^{-1}\sum_{\ell}S_{\ell,I}^{\pd}S_{\ell,J}^{\pd}=1-2\frac{r_{IJ}^{\pd}}{L} .
\end{equation}
This connection between Hamming distance on the FS graph and the spin orientations
will prove important in Sec.\ \ref{subsection:candxbar} in 
relating the real-space spin autocorrelation function (or imbalance) to
the first moment of the FS probabilities.


\section{Diagnosing probability transport}
\label{section:probsintro}

The basic underlying quantities considered are the probabilities 
$P_{IJ}(t)=|G_{IJ}(t)|^{2}\geq 0$, given by
\begin{equation}
\label{eq:PIJdef}
P_{IJ}^{\pd}(t) 
=|\langle J|\Psi(t)\rangle|^{2}
=|\langle J|e^{-i\mathcal{H}t}|I\rangle|^{2}
\end{equation}
with $|\Psi(t)\rangle$ the $t$-dependent wavefunction. 
We add here that, unless stated otherwise, time is shown in units of $\Gamma^{-1}$ in all figures (i.e.\ $\Gamma \equiv 1$).
Physically, $P_{IJ}(t)$  gives the probability that the system will be found on FS site $J$ at time $t$, given its initiation on site $I$ (and with $P_{II}(t)$ the commonly studied return probability). As  reflected in 
$P_{IJ}(t=0) =\delta_{IJ}$, the initial state $|I\rangle$ is  site-localised on the FS graph, and as such wholly Anderson-localised thereon. On increasing $t$, the distribution of probabilities spreads in some fashion through the FS graph/lattice. Understanding at least some aspects of this many-sided process, both temporally and as a function of disorder strength, is the aim of this work.

In the following $\mathcal{H}$ is presumed real symmetric, as relevant to the 
TFI model considered explicitly, such that $P_{IJ}(t)=P_{JI}(t)=P_{IJ}(-t)$.
Expressed in terms of eigenstate amplitudes $A_{nI}=\langle I|n\rangle$,  with
eigenstates $|n\rangle$ and corresponding eigenvalues $E_{n}$, note for later use that
\begin{equation}
\label{eq:PIJamps}
P_{IJ}^{\pd}(t) =
\sum_{n,m} e^{-i(E_{n}^{\pd}-E_{m}^{\pd})t} ~A_{nI}^{\pd}A_{nJ}^{\pd}A_{mI}^{\pd}A_{mJ}^{\pd}.
\end{equation}

 Probability is of course conserved, viz.\ $\sum_{J}P_{IJ}(t) =1$ for all $t$ and any 
initial FS site $I$. For any given $I$, $P_{IJ}(t)$ can thus be regarded as the time-dependent distribution, over all FS sites $J$, of the conserved `mass'  $M_{I}=\sum_{J}P_{IJ}(t) =1$.
A natural way to quantify such a distribution is via its moments. To this end, first  define
\begin{equation}
\label{eq:Prtdef}
\begin{split}
P_{I}^{\pd}(r;t)~=&~\sum_{J:r_{IJ}=r}P_{IJ}^{\pd}(t)
\\
P(r;t)~=~&\nh^{-1}\sum_{I} P_{I}^{\pd}(r;t)
\end{split}
\end{equation}
with the $J$-sum over all $N_{r}=\binom{L}{r}$ FS sites on a given row $r$ of the graph,
for which the Hamming distance  $r_{IJ}=r$ ($I$ lying at the apex of the graph, see 
Fig.\ \ref{fig:fockspace}). $P_{I}(r;t)$ gives the total probability on row $r$ (with 
$\sum_{r=0}^{L}P_{I}(r;t)=1$ $\forall t,I$); its sample average over initial FS sites $I$ is denoted $P(r;t)$. An average over disorder realisations will be denoted, according to convenience,
either by an overbar (e.g.\ $\overline{P}_{I}(r;t)$), or  by angle brackets 
($\langle \cdots\rangle_{\dis}$). The $r$- and $t$-dependence of $\overline{P}(r;t)$ in particular 
will be considered explicitly in Sec.\ \ref{section:verticaltransport}.

Moments of the $\{P_{IJ}(t)\}$ follow directly, e.g.\  the first moment
$r_{I}(t)=\sum_{J}r_{IJ}P_{IJ}(t)= \sum_{r=0}^{L} rP_{I}(r;t)$
and its sample average $r(t)=\nh^{-1}\sum_{I}r_{I}(t)$.
In Sec.\ \ref{section:verticaltransport} we will consider the disorder-averaged moments
\begin{subequations}
\label{eq:rmomentsdef}
\begin{align}
\overline{r}(t)~=&~\sum_{r=0}^{L}~r \overline{P}(r;t)
\label{eq:rbardef}
\\
\overline{\delta r^{2}}(t)=&\sum_{r=0}^{L}~r^{2} \overline{P}(r;t)
-\Big(\sum_{r=0}^{L} r\overline{P}(r;t)\Big)^{2},
\label{eq:dealtar2def}
\end{align}
\end{subequations}
in particular the former. As now shown, for any disorder realisation, $r_{I}(t)$ and $r(t)$ are 
in fact directly related to the real-space spin autocorrelation function; providing thereby 
a direct connection between real-space and Fock-space perspectives.


\subsection{Longitudinal transport and spin autocorrelator}
\label{subsection:candxbar}

Consider $\mathcal{C}(t)$ defined by
\begin{equation}
\label{eq:Cdef}
\mathcal{C}(t)~=~\frac{1}{L}\sum_{\ell=1}^{L}\mathcal{C}_{\ell\ell}^{\pd}(t), ~~~~
\mathcal{C}_{\ell\ell}^{\pd}(t)=\frac{1}{\nh}\mathrm{Tr}(\hat{\sigma}_{\ell}^{z}(t)\hat{\sigma}_{\ell}^{z}),
\end{equation} 
with $\mathcal{C}_{\ell\ell}(t)$ the local real-space spin autocorrelator.
The trace, Tr, can equivalently be either over FS sites,
$\mathcal{C}_{\ell\ell}(t)=\nh^{-1}\sum_{I}\CllI{}(t)$ with 
$\CllI{}(t)=\langle I|\hat{\sigma}_{\ell}^{z}(t)\hat{\sigma}_{\ell}^{z}|I\rangle$,
or over eigenstates, $\mathcal{C}_{\ell\ell}(t)=\nh^{-1}\sum_{n}\Clln{}(t)$.
A simple calculation then relates $\CllI{}(t)$ to the probabilities $\{P_{IJ}(t)\}$,
\begin{equation}
\label{eq:CllI}
C_{\ell\ell}^{[I]}(t)
~=~S_{\ell,I}^{\pd}\sum_{J}S_{\ell,J}^{\pd} P_{IJ}^{\pd}(t)
\end{equation}
where $S_{\ell,I}=\langle I|\hat{\sigma}_{\ell}^{z}|I\rangle$ ($=\pm 1$).
Using Eq.\ \ref{eq:rIJ}, together with conservation of probability, it follows directly that
\begin{equation}
\nonumber
\mathcal{C}^{[I]}(t)~:=~L^{-1}\sum_{\ell}\mathcal{C}_{\ell\ell}^{[I]}(t)
\end{equation}
is given by
\begin{equation}
\label{eq:Cfull}
\begin{split}
C^{[I]}(t)~=~&
1 -\frac{2}{L}\sum_{J} r_{IJ}^{\pd}P_{IJ}^{\pd}(t)
~=~1-\frac{2}{L}r_{I}^{\pd}(t)
\\
\implies ~~
\mathcal{C}(t)~=&~\nh^{-1}\sum_{I}C^{[I]}(t) =1-\frac{2}{L}r(t).
\end{split}
\end{equation}

Eq.\ \ref{eq:Cfull} relates directly the real-space spin autocorrelation function
to the first moment of the FS probabilities $\{ P_{IJ}(t)\}$ (and is not confined to the TFI model, holding equally for XXZ or spinless fermion models).
A striking feature of it  is that,  on timescales for which $\mathcal{C}^{[I]}(t)$
departs by merely a non-vanishing amount from its $t=0$ value of $1$,  the first moment 
$r_{I}(t)\propto L$ is extensive in system size. Intuituively, one expects such timescales to be determined by the hopping energy scale $\Gamma$ which acts to dephase the initially  synchronised spins, and as such to be on the order $\Gamma t \sim \mathcal{O}(1)$. The resultant extensivity 
of $r_{I}(t)$ means that an excitation, initially Anderson-localised on the single FS site $I$, 
spreads significantly throughout the Fock space on the shortest timescales of order 
$\Gamma t \sim \mathcal{O}(1)$ -- and would appear to do so regardless of whether the 
system is ultimately ergodic or MBL. Understanding how this behaviour arises, 
the essential characteristics of the 
Fock-space graph which it reflects,
and the physical picture it gives rise to, is conceptually significant and considered in Sec. \ref{section:shortt} (see also Sec.\ \ref{section:verticaltransport}).

One can also bound $r(t)$. Since $r_{IJ}\leq L$, it follows trivially from
Eq.\ \ref{eq:Cfull} (using $\sum_{J}P_{IJ}(t)=1 ~\forall t$) that $r(t)/L \leq 1$ 
for all $t$. More useful is a bound in the $t\to\infty$ limit. Resolving 
$\mathcal{C}_{\ell\ell}(t)$ as an eigenstate trace, its infinite-time limit 
$\mathcal{C}_{\ell\ell}(\infty)=\nh^{-1}\sum_{n}|\langle n|\hat{\sigma}_{\ell}^{z}|n\rangle|^{2}$, so $\mathcal{C}_{\ell\ell}(\infty)$ and thus $\mathcal{C}(\infty)$ cannot  be negative;  whence 
(Eq.\ \ref{eq:Cfull}) $r(\infty)\leq L/2$ necessarily.
Sufficiently deep in an ergodic phase, with essentially all many-body eigenstates  delocalised and no remnant memory of initial conditions, one expects $\mathcal{C}_{\ell\ell}(\infty)$ to vanish. Hence $r(\infty)=L/2$ -- the midpoint of the FS graph -- is characteristic of such 
`complete' ergodicity. In an MBL phase by contrast, persistent memory of initial conditions means 
$\mathcal{C}_{\ell\ell}(\infty)>0$. In that case, the long-time limit of $r(t)$ is perforce less than $L/2$.


\subsection{Lateral transport}
\label{subsection:lateraltransportA}

For any disorder realisation, $P_{I}(r;t)$ gives (Eq.\ \ref{eq:Prtdef}) the 
total probability on row $r$ of the graph/lattice. Study of  its $(r,t)$-dependence thus reveals how probability flows in time `down' the FS graph, row by row. It does not however give information on the important issue of how the distribution of probabilities spreads out laterally, and in general inhomogeneously, across the rows of the graph.

One such measure of the latter, studied numerically in 
Sec.\ \ref{section:lateraltransportmain},  is provided by $R_{I}(r;t)\geq 1$ defined by
\begin{equation}
\label{eq:RIdef}
R_{I}^{\pd}(r;t)~=~\frac{\frac{1}{N_{r}}\underset{J:r_{IJ}=r}{\sum} P^{2}_{IJ}(t)}{\big(\frac{1}{N_{r}} \underset{J:r_{IJ}=r}{\sum}P_{IJ}(t)\big)^{2}} .
\end{equation}
For any \emph{given} disorder realisation, this is simply the ratio of the mean squared 
probability per FS site on row $r$, to the square of the corresponding mean probability, 
$[N_{r}^{-1}P_{I}(r;t)]^{2}$. So it provides an obvious measure of fluctuations in the distribution of $P_{IJ}$'s along a given row. In particular, $R_{I}(r;t)= 1$ in a limit 
of extreme homogeneity where all $P_{IJ}(t)$'s on the row are the same.
The latter behaviour will in fact be shown in Sec.\ \ref{section:shortt} to arise at
sufficiently short times, \emph{independently} of disorder strength $W$; before evolving in 
$t$ to a distribution which is $W$-dependent, and strongly inhomogeneous in the MBL regime
(Sec.\ \ref{section:lateraltransportmain}).

The average of $R_{I}(r;t)$ over disorder realisations and FS sites $I$ will be denoted for brevity
by $\langle R\rangle  \equiv \langle R\rangle(r;t)$,
\begin{equation}
\label{eq:Rmeandef}
\langle R\rangle ~=~\nh^{-1}\sum_{I}\big\langle R_{I}^{\pd}(r;t)\big\rangle_{\dis}^{\pd} 
\end{equation}
with $\langle \cdots\rangle_{\dis}$ the disorder average. More generally, we also study in 
Sec.\ \ref{subsection:vertprobdists} the full probability distribution of $R_{I}(r;t)$, given by
\begin{equation}
\label{eq:PdistR}
P_{R}^{\pd}(x) ~=~\nh^{-1}\sum_{I}\big\langle \delta\big(x-R_{I}^{\pd}(r;t)\big)\big\rangle_{\dis}^{\pd} 
\end{equation}
(of which the first moment is $\int dx~xP_{R}(x)=\langle R\rangle$).
In the MBL regime in particular, $P_{R}(x)$ at sufficiently long times will be shown
to be characterised by a heavy-tailed L\'evy alpha-stable distribution.

The quantity $R_{I}(r;t)$ is directly related to another natural measure of fluctuations
in the distribution of $P_{IJ}(t)$'s along a given FS row: the row-resolved,  
$t$-dependent inverse participation ratio (IPR). To motivate this, consider the $t$-dependent wavefunction following the initial quench, $|\Psi(t)\rangle =e^{-i\mathcal{H}t}|I\rangle$, 
expanded as $|\Psi(t)\rangle =\sum_{J}\AJI{J}(t)|J\rangle$; such that, from 
Eq.\ \ref{eq:PIJdef}, the squared amplitudes $|\AJI{J}(t)|^{2}=P_{IJ}(t)$ are just the probabilities of interest. Time-dependent wavefunction densities, normalised on any given row 
$r$, are then given by $|\BJI{J}(t)|^{2}=|\AJI{J}(t)|^{2}/\sum_{J:r}|\AJI{J}(t)|^{2}$ (with 
$\sum_{J:r}$ shorthand  for $\sum_{J:r_{IJ}=r}$); for which the associated generalised IPR is
$\mathcal{I}_{I,q}=\sum_{J:r}|\BJI{J}(t)|^{2q}$. Hence, for the standard case of $q=2$ on 
which we focus explicitly, the IPR is related simply to $R_{I}(r;t)$ (Eq.\ \ref{eq:RIdef})
 via $N_{r}=\binom{L}{r}$,
\begin{equation}
\label{eq:IPRandR}
\begin{split}
\mathcal{I}_{I,2}^{\pd}(r;t)~=&~
\frac{\underset{J:r_{IJ}=r}{\sum} P^{2}_{IJ}(t)}{\big(\underset{J:r_{IJ}=r}{\sum}P_{IJ}(t)\big)^{2}}
~=~N_{r}^{-1}R_{I}^{\pd}(r;t)
\\
\implies 
\langle \mathcal{I}_{2}^{\pd}\rangle ~=~&\nh^{-1}\sum_{I}\big\langle \mathcal{I}_{I,2}^{\pd}(r;t)\big\rangle_{\dis}^{\pd}
~=~N_{r}^{-1}\langle R\rangle 
\end{split}
\end{equation}
(with the corresponding probability distribution of $\mathcal{I}_{I,2}$ following
trivially from that for $R_{I}(r;t)$, Eq.\ \ref{eq:PdistR}).

We can then reason physically as follows, considering some particular time $t$.
If the amplitudes $|\BJI{J}(t)|^{2}=P_{IJ}(t)/\sum_{J:r}P_{IJ}(t)$ -- and hence the 
probabilities $P_{IJ}(t)$ -- are essentially uniformly distributed over the $N_{r}$ 
FS sites on row $r$, then each $|\BJI{J}|^{2}\sim N_{r}^{-1}$. Hence 
$\langle\mathcal{I}_{2}\rangle \sim N_{r}^{-1}$ and in turn
$\langle R\rangle \sim \mathcal{O}(1)$ should be of order unity (and as such $L$-independent).
If by contrast the wavefunction is strongly inhomogeneously distributed on the row, one
might anticipate $\langle\mathcal{I}_{2}\rangle \sim N_{r}^{-\nu}$ with a fractal exponent 
$\nu \equiv \nu(t)<1$; and hence from Eq.\ \ref{eq:IPRandR} that 
$\langle R\rangle \sim N_{r}^{1-\nu}$ -- which thus grows with increasing system size $L$.
These two behaviours will indeed be shown to arise in Sec.\  \ref{section:lateraltransportmain}, the former characteristic at long time of the ergodic regime, and the latter characteristic of the MBL regime at larger disorder strengths.

Finally, as a complement to $P_{R}(x)$ (Eq.\ \ref{eq:PdistR}), we also study in 
Sec.\  \ref{subsection:vertprobdists} the probability distribution
\begin{equation}
\label{eq:Preldefn}
P_{\rel}^{\pd}(x) =
\frac{1}{\nh}\sum_{I}\frac{1}{N_{r}}\underset{J:r_{IJ}=r}{\sum}\Big\langle
\delta\Big(x -\frac{P_{IJ}^{\pd}(t)}{\frac{1}{N_{r}}\underset{J:r_{IJ}=r}{\sum}P_{IJ}^{\pd}(t)}\Big)
\Big\rangle_{\dis}^{\pd}.
\end{equation}
For any given row $r$ on the graph, this gives the distribution of $P_{IJ}(t)$ relative to its mean value on the row, $x =P_{IJ}(t)/[N_{r}^{-1}P_{I}(r;t)]$; its second moment being precisely
$\int dx~x^{2}P_{\rel}(x)=\langle R\rangle$, see Eqs.\ \ref{eq:Rmeandef},\ref{eq:RIdef} (and 
its first is $1$ by construction).

\section{Short-time behaviour}
\label{section:shortt}

We turn now to the short-time behaviour of probability transport for the disordered TFI model.
While the underlying calculations are simple, the physical picture arising is rather rich;
including the emergence at short times of multifractality in the $t$-dependent wavefunction
$|\Psi(t)\rangle =e^{-i\mathcal{H}t}|I\rangle$ -- for \emph{any} disorder strength $W$, and 
as such independent of whether the  ultimate long-time behaviour of the system is ergodic 
or MBL in nature.

Consider $P_{IJ}(t)=|G_{IJ}(t)|^{2}$, where (Eq.\ \ref{eq:PIJdef})
\begin{equation}
\label{eq:GIJexp}
G_{IJ}^{\pd}(t) ~=~\langle J|e^{-i\mathcal{H}t}|I\rangle ~=~\sum_{n=0}^{\infty}\frac{(-i)^{n}}{n!}t^{n}
\langle J| \mathcal{H}^{n}|I\rangle,
\end{equation}
and  separate $\mathcal{H}\equiv \mathcal{H}_{0}+\mathcal{H}_{1}$ (Eq.\ \ref{eq:hamTBM}), with 
$\mathcal{H}_{0}=\sum_{K}\mathcal{E}_{K}|K\rangle\langle K|$ and $\mathcal{H}_{1}$ the hopping term.
With $K_{m}$ denoting any FS site on row $m$, $\mathcal{H}|K_{m}\rangle$ connects solely to FS sites in rows $m\pm 1$ (and $m$),  since non-zero hopping matrix elements ($\Gamma$) connect only FS sites on adjacent rows of the graph.
Now let $J$ in Eq.\ \ref{eq:GIJexp} be some given FS site on row $r$, call it $J_{r}$. Obviously, 
 $\langle J_{r}|\mathcal{H}^{n}|I\rangle$ vanishes identically for all $n<r$. Hence
\begin{equation}
\label{eq:GIJshortt}
G_{IJ_{r}^{\pd}}(t) ~=~\frac{(-i)^{r}}{r!}t^{r}\langle J_{r}^{\pd}| (\mathcal{H}_{1})^{r}|I\rangle ~+~\mathcal{O}(t^{r+1}) .
\end{equation}
The leading term here will clearly dominate $G_{IJ_{r}}(t)$ for sufficiently small $t$. Importantly, it involves \emph{solely} FS hoppings, consisting of `forward paths' from $I$ 
to $J_{r}$, each containing precisely $r$ hops (i.e.\ $r$ factors of $\Gamma$). For any given FS site $J_{r}$ there are however $r!$ identical contributions to 
$\langle J_{r}| (\mathcal{H}_{1})^{n}|I\rangle$, because there are $r!$ distinct forward paths from $I$ to $J_{r}$ on the FS graph; and each such contribution has a value of $\Gamma^{r}$.
This cancels the $1/r!$ factor in Eq.\ \ref{eq:GIJexp}, from which the leading small-$t$ behaviour is $G_{IJ_{r}^{\pd}}(t)\sim (-i)^{r} (\Gamma t)^{r}$, and that of $P_{IJ_{r}}(t)$ thus
\begin{equation}
\label{eq:PIJshortt}
P_{IJ_{r}^{\pd}}(t)~\sim ~ (\Gamma t)^{2r}  .
\end{equation}
Note the following points about this leading short-time behaviour:\\
(i) It holds for any $r$, and for all FS sites on row $r$. By virtue of the latter, the distribution of probabilities along any given row is fully homogeneous in the time window over which Eq.\ \ref{eq:PIJshortt} holds. In consequence,  $R_{I}(r;t)=1$ (Eq.\ \ref{eq:RIdef}), 
the distribution $P_{R}(x)=\delta(x-1)$ (Eq.\ \ref{eq:PdistR}) is $\delta$-distributed, and
the row-resolved IPR $\mathcal{I}_{2}(r;t)=N_{r}^{-1}$ (Eq.\ \ref{eq:IPRandR}).
By itself the above calculation does not of course prescribe the timescale over which such behaviour occurs, but we ascertain it below. (ii) Relatedly, since solely the 
disorder-independent hoppings $\Gamma$ generate Eq.\ \ref{eq:PIJshortt}, the result is 
independent of  disorder strength $W$. (iii) Although $P_{IJ_{r}}(t)\sim (\Gamma t)^{2r}$ decreases exponentially rapidly with $r$, the number $N_{r}=\binom{L}{r}$ of FS sites on row $r$ grows exponentially with $r$. Hence, even for short times, one cannot neglect the contribution of  sites on any row $r$ to e.g.\ the first moment of the probability distribution,
$r_{I}(t)=\sum_{J}r_{IJ}P_{IJ}(t)\equiv \sum_{r=0}^{L}\binom{L}{r} r P_{IJ_{r}}(t)$,
as considered below.
(iv) The calculation above naturally reflects the intrinsic structure of the FS
graph (Fig.\ \ref{fig:fockspace}) for the disordered TFI model. We simply remark that
the result arising would be quite different if one considered a tree graph (Cayley tree/Bethe lattice); for while in that case Eq.\ \ref{eq:GIJshortt} holds for any given site $J_{r}$ on generation $r$ of the tree, there is just a single  path connecting the root site $I$ to the given $J_{r}$.

As it stands, direct use of Eq.\ \ref{eq:PIJshortt} for each $r$ fails to conserve total probability. This however is readily taken into account by writing 
$P_{IJ_{r}}(t)=g(r;t) (\Gamma t)^{2r}$ where, for all $r$, $g(r;t)$ must satisfy
 (a) $g(r;t$$=$$0) =1$, such that the leading low-$t$ behaviour of $P_{IJ_{r}}(t)$ is 
Eq.\ \ref{eq:PIJshortt}; (b) $g(r;t) >0$ for all times for which the calculation is valid;
and (c) overall probability must be conserved, $\sum_{J}P_{IJ}(t)=1$, i.e.\
$\sum_{r=0}^{L}\binom{L}{r} g(r;t) (\Gamma t)^{2r}=1$ $\forall t$. This has the solution 
$g(r;t)=[ 1-(\Gamma t)^{2}]^{(L-r)}$. And $g(r;t) >0$ $\forall r$ is satisfied provided 
$\Gamma t <1$, which upper bounds the time-window over which the calculation holds.

 \begin{figure}
\includegraphics[width=\columnwidth]{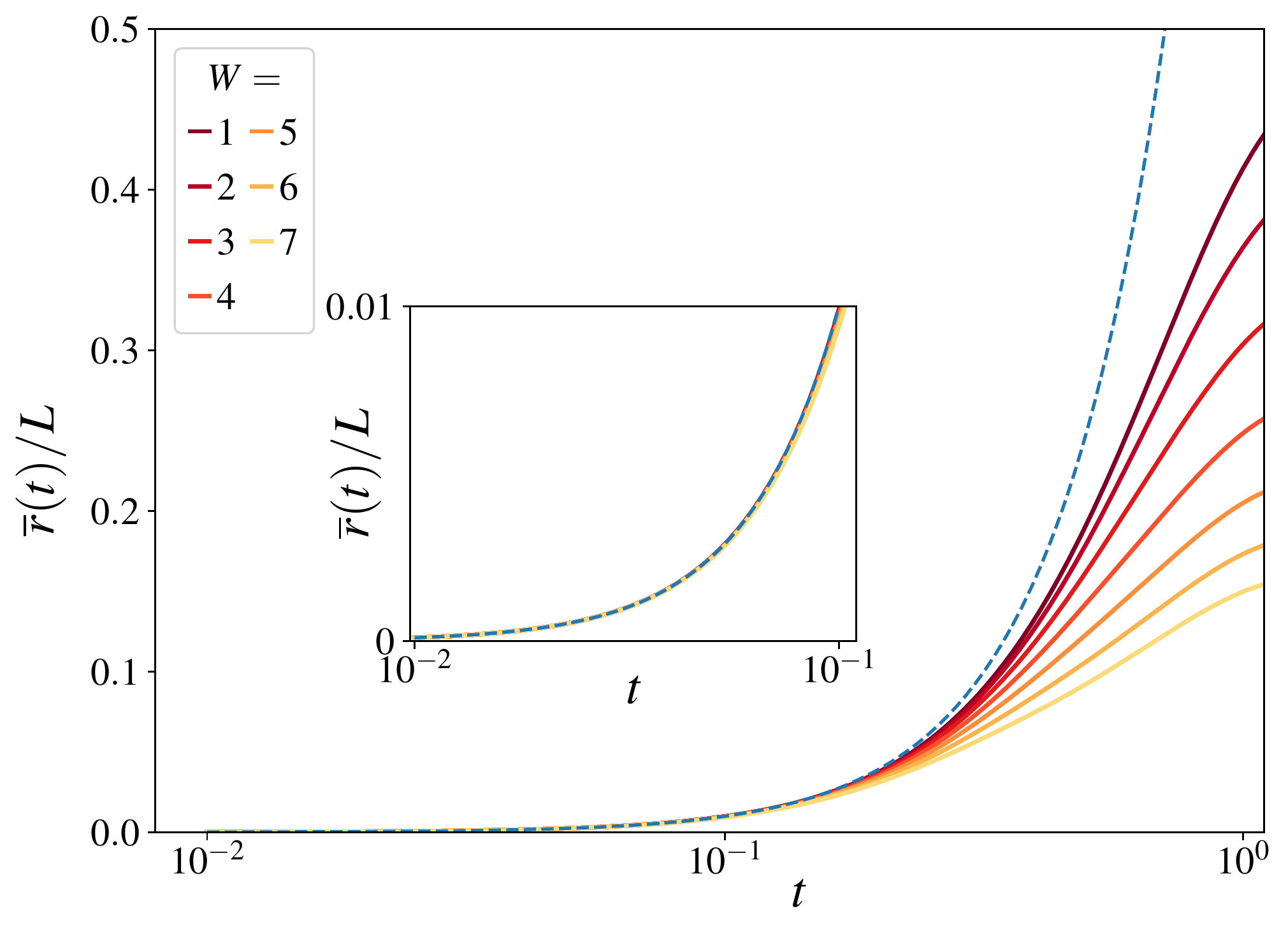}
   \caption{ ED results for $\bar{r}(t)/L$ \emph{vs} $t$ ($\Gamma \equiv 1$), shown
	over the indicated range of disorder strengths $W$.
	For times $ t \lesssim 0.1$, the $W$-independent behaviour of
	Eq.\ \ref{eq:rbarshorttfull} (dashed line) is seen to arise.
	Inset: Same results, on smaller $ t$-scale.
	} 
	\label{fig:small_t_rbar}
\end{figure}

The essential result for the short-$t$ behaviour of $P_{IJ_{r}}(t)$ is then
\begin{equation}
\label{eq:PIJshorttfull}
P_{IJ_{r}^{\pd}}^{\pd}(t) ~= ~ \big(\Gamma t\big)^{2r} \big[ 1-(\Gamma t)^{2}\big]^{(L-r)} .
\end{equation}
As this is independent of both the initial FS site $I$ and disorder strength $W$, 
the resultant first moments (Sec.\ \ref {section:probsintro})
$r_{I}(t)\equiv r(t)\equiv \overline{r}(t)$ coincide, and follow from Eq.\ \ref{eq:PIJshorttfull} 
as
\begin{equation}
\label{eq:rbarshorttfull}
r_{I}^{\pd}(t)~\equiv~\overline{r}(t) ~=~\sum_{r=0}^{L}\binom{L}{r} ~r ~P_{IJ_{r}^{\pd}}^{\pd}(t) ~=~L (\Gamma t)^{2}.
\end{equation}
That the short-time behaviour is $W$-independent is corroborated in Fig.\ \ref{fig:small_t_rbar}, which shows ED results for $\overline{r}(t)/L$ \emph{vs} $\Gamma t$, over a range of disorder strengths $W$. In all cases, the asymptotic behaviour Eq.\ \ref{eq:rbarshorttfull} 
indeed arises at short times -- in practice for $\Gamma t \lesssim 0.1$ or so, consistent with the bound above.

The fact that $\overline{r}(t) \propto L$ is extensive for finite $\Gamma t$ means of 
course that it is $\overline{r}(t)/L$ which remains finite in the thermodynamic limit 
$L\to \infty$. The relevant fluctuations in this quantity are thus embodied in
$\overline{\delta r^{2}}(t)/L^{2}$,  direct evaluation of which using Eq.\ \ref{eq:PIJshorttfull} gives  $\overline{\delta r^{2}}(t)/L^{2}=(\Gamma t)^{2}[1-(\Gamma t)^{2}]/L$. Since this is 
$\propto 1/L$, such fluctuations vanish in the thermodynamic limit, with $r(t)/L$ 
$\delta$-distributed on its mean.

Finally, although by itself a somewhat limited diagnostic of probability transport on Fock space, we comment parenthetically on the commonly studied~\cite{SantosPRB2015,SantosPRBRC2018,SchiulazPRB2019}
return probability, $P_{II}(t)$. This corresponds to $r=0$ in Eq.\ \ref{eq:PIJshorttfull}, which 
for $\Gamma t \ll 1$ recovers the known behaviour~\cite{SchiulazPRB2019} 
$P_{II}(t)\sim \exp(-L(\Gamma t)^{2})$, whereby for any non-zero $\Gamma t$, even if small, the return probability is exponentially suppressed in system size $L$.

\subsection*{Emergent multifractality}
\label{subsection:shorttmultif}

As shown above, for $\Gamma t$ small compared to unity but finite, the probability density has  spread through Fock space to macroscopically large Hamming distances on the order of $L$.
The probabilities $P_{IJ_{r}}(t)$  are uniform on any given row of the FS graph  
(Eq.\ \ref{eq:PIJshorttfull}),  symptomatic of which the row-resolved IPR (Eq.\ \ref{eq:IPRandR}) is  $\mathcal{I}_{I,2}(r;t)=N_{r}^{-1}$.

One can however  also ask for the behaviour of the conventional IPR over the \emph{full} 
Fock-space. For a wavefunction $|\Psi(t)\rangle =\sum_{J}\AJI{J}(t)|J\rangle$, with 
squared amplitudes $|\AJI{J}(t)|^{2}$ ($=P_{IJ}(t)$) normalised to unity over 
all FS sites $J$, the generalised ($q$-dependent) IPR is defined by
$\mathcal{L}_{I,q}(t) =\sum_{J} |\AJI{J}(t)|^{2q}=\sum_{J}P_{IJ}^{q}(t)$;
where only $q>1$ is considered henceforth (trivially, for all $t$, $\mathcal{L}_{I,0}(t)=\nh$ 
and  $\mathcal{L}_{I,1}(t)=1$). The $L$-dependence of $\mathcal{L}_{I,q}^{\pd}(t)$ is
embodied in the exponent $\tau_{q}\equiv \tau_{q}(t)$ defined by 
$\mathcal{L}_{I,q}(t)~\sim~\nh^{-\tau_{q}}$. If $\tau_{q}=0$  for any specified $t$, 
then the wavefunction $|\Psi(t)\rangle$ is Anderson localised on $\mathcal{O}(1)$  FS sites 
of the graph/lattice, while if $\tau_{q}=q-1$ it is essentially uniformly spread over all FS sites on the graph, and as such ergodic. But if by contrast $0<\tau_{q}<q-1$, then the wavefunction is fractal; more specifically,  if $\tau_{q}$ is a non-linear function of $q$, then it is multifractal.

To consider this in the present context, it is convenient to rewrite Eq.\ \ref{eq:PIJshorttfull}
in the binomial form
\begin{equation}
\label{eq:PIJrbinom}
P_{IJ_{r}^{\pd}}^{\pd}(t) ~=~[z(t)]^{r}[1-z(t)]^{(L-r)}
\end{equation}
with $z(t)=(\Gamma t)^{2}$ for short times $\Gamma t \ll 1$. This in turn can be expressed as
\begin{equation}
\label{eq:PIJrshorttXi}
P_{IJ_{r}^{\pd}}^{\pd}(t) ~=~\big[1+e^{-1/\xi_{F}(t)}\big]^{-L} e^{-r/\xi_{F}(t)},
\end{equation}
in terms of a correlation length $\xi_{F}(t)$ defined by
$\xi_{F}^{-1}(t) = \ln(\tfrac{1}{z(t)}-1)$. Since the short-time $P_{IJ_{r}}(t)$'s are the 
same for all $\binom{L}{r}$ sites on row $r$ of the graph, 
$\mathcal{L}_{I,q}(t)\equiv \sum_{r=0}^{L}\binom{L}{r} P_{IJ_{r}}^{q}(t)$. Hence from 
Eq.\ \ref{eq:PIJrshorttXi}
\begin{equation}
\label{eq:tauqshortt}
\tau_{q}^{\pd}(t) ~=~ \log_{2}\left[\frac{\big(1+e^{-1/\xi_{F}(t)}\big)^{q}}{\big(1+e^{-q/\xi_{F}(t)}\big)}
\right],
\end{equation}
where $e^{-1/\xi_{F}(t)} \sim (\Gamma t)^{2}$ for $\Gamma t \ll 1$.
For $t=0$ precisely, $\tau_{q}=0$. This is just as expected, reflecting the fact that 
$|\Psi(t$$=$$0)\rangle =|I\rangle$ is Anderson localised on the FS graph.

However, for any non-zero $\Gamma t\ll 1$, Eq.\ \ref{eq:tauqshortt} is readily seen to be
non-linear in $q$ and to satisfy $0<\tau_{q}(t) \ll q-1$.  The wavefunction is thus multifractal.
Moreover, this behaviour arises for any disorder strength $W$.
 Emergent multifractality at short times is therefore common both to $W$'s for which
the system is ergodic in the long-time limit, as well as for $W$'s for which it is MBL at long times. In the latter case, one anticipates continued persistence of multifractality beyond the short-time window. In the ergodic case by contrast, one expects multifractality to dissipate with further increasing $t$, as the distribution of probabilities homogenises over the entire
graph and  the long-time limit of $\tau_{q}(t=\infty)=q-1$ arises~\footnote{That $\tau_{q}(t$$=$$\infty)=q-1$  in this case  is easily seen from an argument applicable deep in an ergodic phase, where eigenstates are effectively Gaussian  random vectors (grv's). From Eq.\ \ref{eq:PIJamps}, $P_{IJ}(\infty)=\sum_{n}A_{nI}^{2}A_{nJ}^{2}$ quite generally, with $A_{nI}^{2} \sim 1/\nh$ for the case of grv's. Hence $P_{IJ}(\infty)\sim 1/\nh$, from which $\mathcal{L}_{I,q}(\infty)=\sum_{J}P_{IJ}^{q}(\infty)$ $\sim\nh^{1-q}=\nh^{-\tau_{q}(\infty)}$.}.

\begin{figure}
\includegraphics[width=\columnwidth]{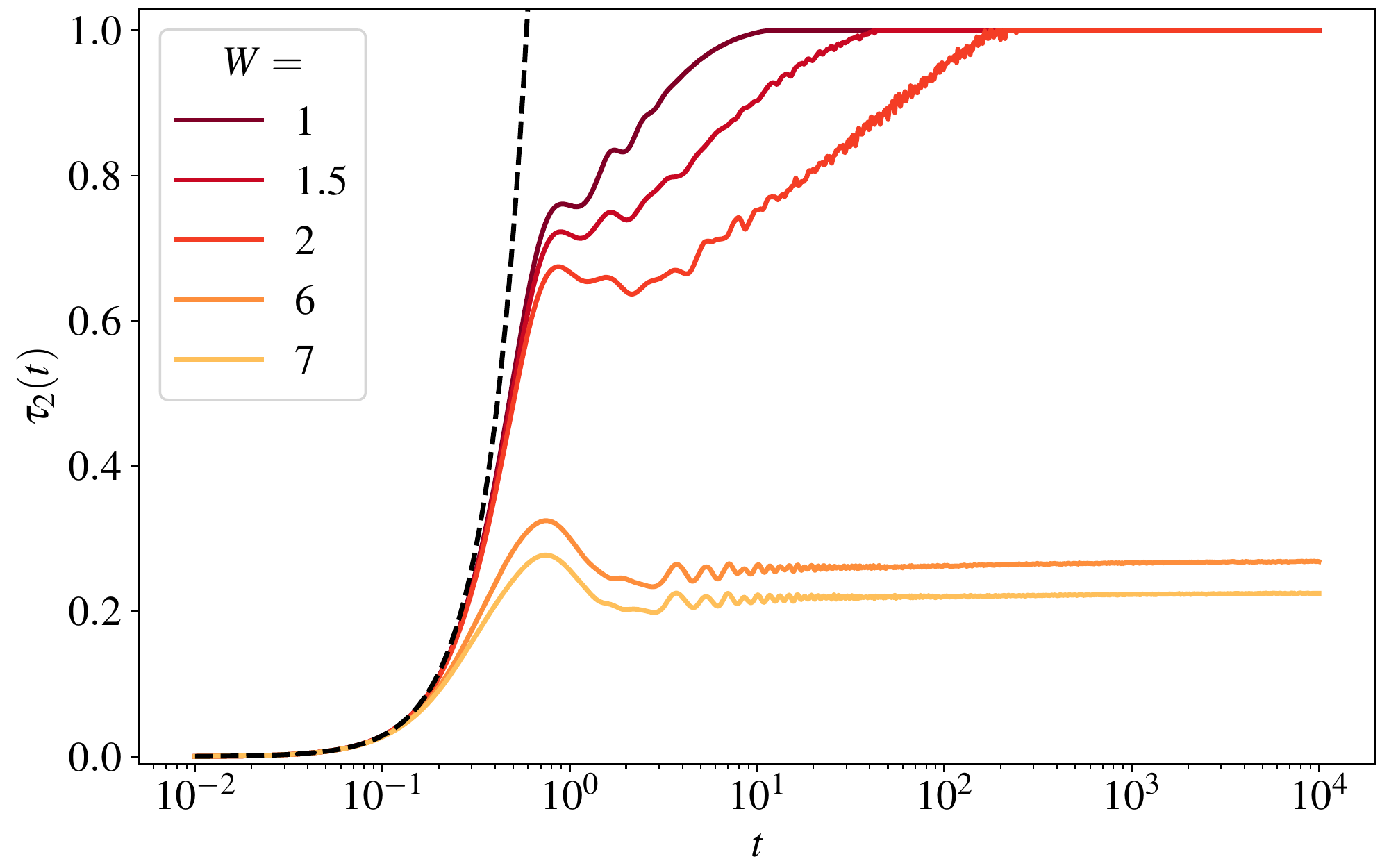}
  \caption{
	 $t$-dependence of the IPR exponent $\tau_{2}(t)$ obtained from ED calculations, with $W=1,1.5,2$
	exemplifying the ergodic phase and $W=6,7$ the MBL regime.
	Dashed line shows the low-$t$ asymptotic behaviour $\tau_{2}(t)\sim 2(\Gamma t)^{2}/\ln2$
	(from Eq.\ \ref{eq:tauqshortt}). Full discussion in text.
	} 
	\label{fig:tau2t-dep}
\end{figure} 

That the above behaviour indeed arises is illustrated in Fig.\ \ref{fig:tau2t-dep} which, for the standard case $q=2$, shows ED results for the $t$-dependent exponent $\tau_{2}(t)$.
We define the latter in general via the averaged IPR, 
$\overline{\mathcal{L}}_{2}(t) =\nh^{-1}\sum_{I}\overline{\mathcal{L}}_{I,2}(t)$,
written as $\overline{\mathcal{L}}_{2}(t)=c(t)\nh^{-\tau_{2}(t)}$ (adding that for short times 
$\Gamma t\ll 1$, this definition is the same as that arising from Eq.\ \ref{eq:tauqshortt}, 
since $P_{IJ_{r}}(t)$ in Eq.\ \ref{eq:PIJrshorttXi} is independent of both disorder and
the FS site $I$). For any chosen $t$, a plot of $\ln \overline{\mathcal{L}}_{2}(t) $ \emph{vs} 
$\ln \nh \propto L$ then gives $-\tau_{2}(t)$ from the slope ($c(t)$ is assumed to be
$L$-independent); and very good linear fits are indeed found for the data shown.

As seen in Fig.\ \ref{fig:tau2t-dep}, for short times $\Gamma t \lesssim 0.1$, the
$W$-independent result from Eq.\ \ref{eq:tauqshortt} is indeed recovered,
viz.\ $\tau_{2}(t)\sim 2(\Gamma t)^{2}/\ln2$, and the wavefunction is multifractal for
all $W$. For $W=6,7$ illustrative of the MBL regime, $\tau_{2}(t)$ remains $<1$ 
and multifractality persists. 
But for $W=1,1.5,2$ illustrating the ergodic regime,
$\tau_{2}(t)$ grows with increasing $t$ and ultimately plateaus to
a long-time value of $\tau_{2}=1$ ($\equiv q-1$), indicating ergodic behaviour.

\section{Longitudinal probability transport }
\label{section:verticaltransport}

\begin{figure}
\includegraphics[width=\columnwidth]{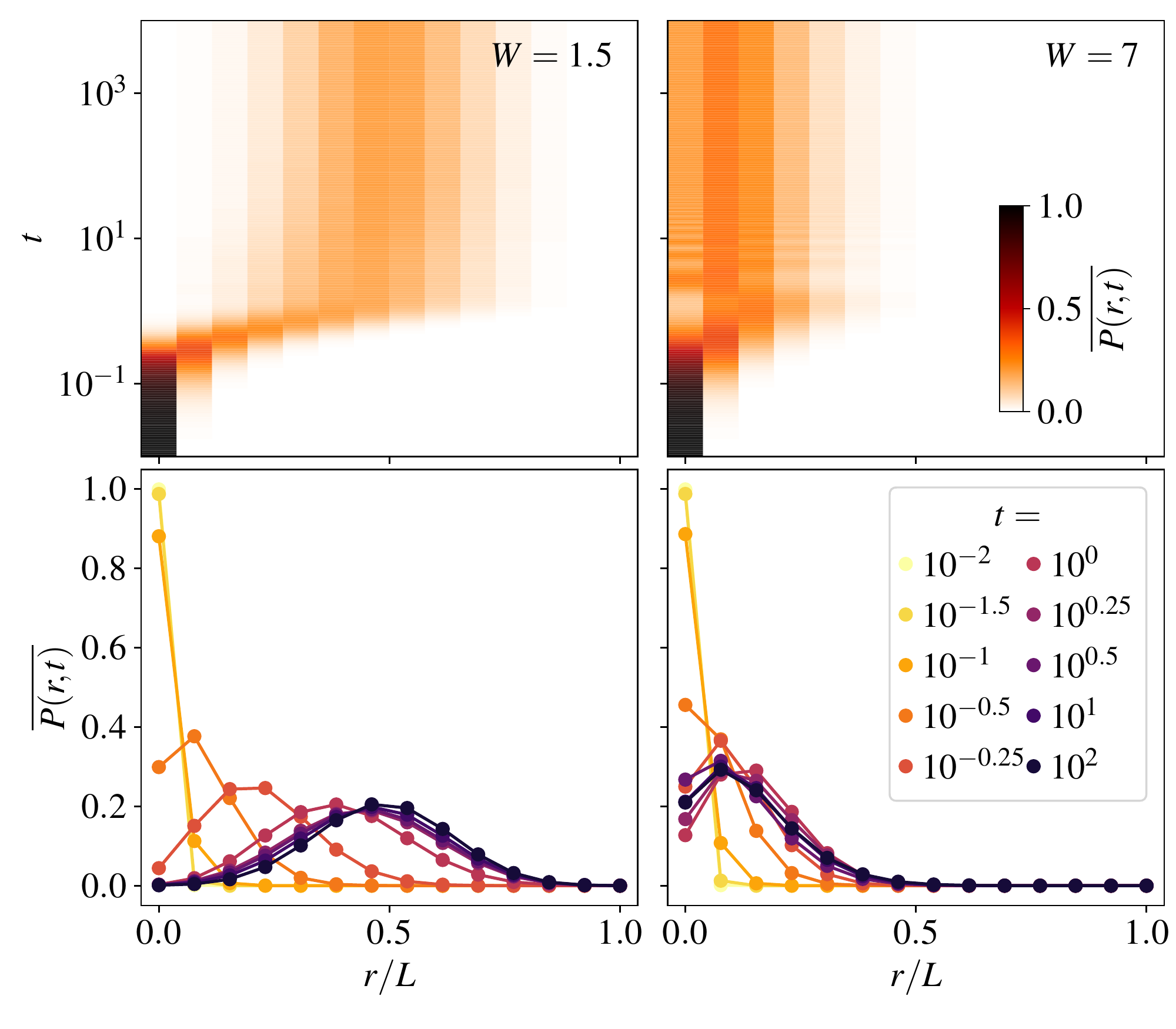}
  \caption{
	 $t$-dependent Fock-space distribution $\overline{P}(r;t)$, for W=1.5 (ergodic phase) in the left column, and for W=7 (MBL) in the right column. Top panels show 
	$\overline{P}(r;t)$  as a colour map in the $(r,t)$-plane, with white denoting $0$ and black denoting $1$. Bottom panels show $\overline{P}(r;t)$ as a function of $r/L$ for different time slices as indicated in the legend.
Data for $L=14$, averaged over $2-3\times 10^{3}$ disorder realisations.  
	} 
	\label{fig:Pbarrtmap}
\end{figure}

In this section we consider how, following a $t=0$ quench into some FS site, probability flows in time down the FS graph, row by row.

To give an initial broad overview, Fig.\ \ref{fig:Pbarrtmap} shows the $r$- and $t$-dependence of the disorder-averaged total probability on row $r$, $\overline{P}(r;t)$ 
(Eq.\ \ref{eq:Prtdef}); for $W=1.5$ (left panels) as representative of the ergodic phase,  and for $W=7$ (right panels) as typical of the MBL regime. The top panels show  $\overline{P}(r;t)$  as a colour map in the $(r,t)$-plane,  while the bottom panels show it as function of $r/L$, for the (logarithmic) sequence of time slices indicated. The qualitative features arising are clear. At short times, $\Gamma t\lesssim 0.1$,  $\overline{P}(r;t)$ is the same for both $W$'s, as expected from the considerations of Sec.\ \ref{section:shortt}.
The distributions begin to spread out in an obvious sense for times $\Gamma t \gtrsim 0.5$, and in practice reach their long-time steady state  by $\Gamma t \sim 10^{1} -10^{2}$. 
For the $W=1.5$ example, the mode of the long-time $\overline{P}(r;t)$  lies at $r/L=1/2$, 
the mid-point of the FS graph; and its $r$-profile is Gaussian (with a width that decreases with
increasing system size, a point to which we return later).
Similar behaviour is found for $W=7$, but with the notable difference that in this
case the mode of the long-time $\overline{P}(r;t)$ occurs at an $r/L$ that is markedly
less than $1/2$.

Quite a bit of information is contained in plots such as Fig.\ \ref{fig:Pbarrtmap}.
To interrogate it, we turn now to the first moment of the probability distribution,
$\overline{r}(t)=\sum_{r=0}^{L}r\overline{P}(r;t)$ (Eq.\ \ref{eq:rbardef}).
More specifically, we consider $\overline{r}(t)/L$, since it is this quantity which necessarily remains finite in the thermodynamic limit $L\to \infty$ (Secs.\ \ref{section:probsintro},\ref{section:shortt}). We add here that in all figures shown in the paper, disorder averages are taken over a minimum of $10^{3}$ realisations.


\subsection{$\boldsymbol{\overline{r}(t)}$: ergodic regime}

\label{subsection:rbar(t)}

 \begin{figure}
\includegraphics[width=\columnwidth]{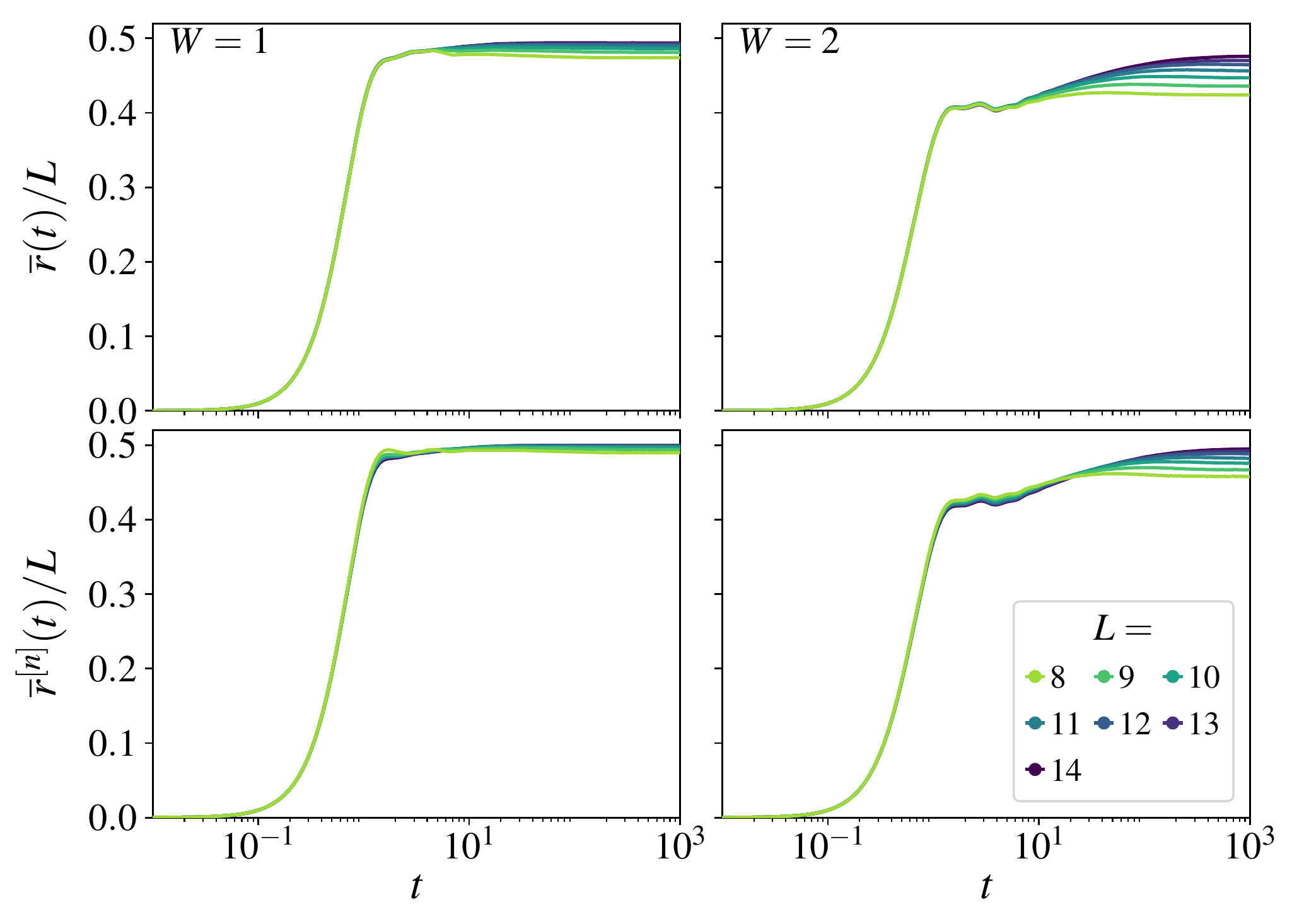}
  \caption{For $W=1$ and $2$, upper panels show ED results for $\bar{r}(t)/L$  \emph{vs} $t$, 
	for $L=8-14$.
	Lower panels show corresponding $\barrn{}(t)/L$ for band centre eigenstates $n$.
	Full discussion in text.
	} 
	\label{fig:rbar(n)W=1,2}
\end{figure}

For disorder strengths $W=1,2$, the upper panels in Fig.\ \ref{fig:rbar(n)W=1,2}
show the $t$-dependence of $\overline{r}(t)/L$, over a time-window comparable to or in excess of the associated Heisenberg times $\tH$ (the inverse of the mean level spacing, $\tH$ is discussed briefly in 
Appx.\ \ref{section:Heistimes} and given for the model by Eq.\ \ref{eq:AppHeis1}).
For both $W$'s, $\overline{r}(t)/L$ for $\Gamma t\lesssim 0.1$ is given by the $W$- and 
$L$-independent short-time result Eq.\ \ref{eq:rbarshorttfull} (as shown in 
Fig.\ \ref{fig:small_t_rbar}). For the $W=1$ case, on further increasing $t$,  
$\overline{r}(t)/L$ rapidly increases towards a value which, for practical purposes,
is $\sim 1/2$ for $\Gamma t\gtrsim 10$ or so. For $W=2$, the situation is rather different. In that case, while $\overline{r}(t)/L$ again grows rapidly up to around $\Gamma t \sim 1$, 
the  `elbow' seen in Fig.\ \ref{fig:rbar(n)W=1,2} around this time is succeeded at longer, intermediate times by a regime of slower dynamics, and the long-time limit is discernibly $<1/2$ for the largest system size studied.

To obtain some understanding here, it is first helpful to consider the infinite-$t$ limit.
From Eq.\ \ref{eq:PIJamps}, $P_{IJ}(\infty)=\sum_{n}A_{nI}^{2}A_{nJ}^{2}$, from which
$\overline{r}(\infty)=\nh^{-1}\sum_{I,J} r_{IJ}\overline{P_{IJ}}(\infty)$ can be expressed as
\begin{subequations}
\begin{align}
\overline{r}(\infty) ~=&~\nh^{-1}\sum_{n}\sum_{r=0}^{L} r \overline{F}_{n}^{\pd}(r)
\label{eq:rbarinftF}
\\
\mathrm{with}~~~
\overline{F}_{n}^{\pd}(r)~=&~\sum_{I,J: r_{IJ}=r} \overline{A_{nI}^{2}A_{nJ}^{2}}.
\label{eq:Fnbardef}
\end{align}
\end{subequations}
$\overline{F}_{n}(r)$ itself was studied in detail in~[\onlinecite{roy2021fockspace}],
where it was shown to be of form
\begin{equation}
\label{eq:Fnbarinf}
\overline{F}_{n}^{\pd}(r)~=~\binom{L}{r} \big(1+e^{-1/\xi_{F,n}^{\pd}}\big)^{-L} e^{-r/\xi_{F,n}^{\pd}},
\end{equation}
with $\xi_{F,n}$ a FS correlation length for eigenstates $n$ at the particular energy 
$\w$ considered  (while band centre states $\w =0$ were considered explicitly 
in~[\onlinecite{roy2021fockspace}], there is nothing special about this energy).
Eqs.\ \ref{eq:rbarinftF},\ref{eq:Fnbarinf} give
\begin{equation}
\label{eq:rbarinftsum}
\frac{\overline{r}(\infty)}{L} =\nh^{-1}\sum_{n} \frac{1}{1+e^{1/\xi_{F,n}^{\pd}}}
=\int d\w 
\frac{D(\w)}{1+e^{1/\xi_{F}^{\pd}(\w)}}
\end{equation}
with $D(\w)=\nh^{-1}\sum_{n}\delta(\w -E_{n})$ the (self-averaging) 
many-body density of states.
The behaviour of $\xi_{F}(\w)$ with disorder strength $W$ is known from a detailed 
scaling analysis~[\onlinecite{roy2021fockspace}]. For $W$'s greater than the critical 
$\wc(\w)$ for which states at the chosen energy $\w$  become MBL, $\xi_{F}(\w)$ remains 
finite (including $W=\wc(\w)^{+}$). For $W<\wc(\w)$ by contrast, $\xi_{F}(\w)\propto L$ and 
thus diverges in the thermodynamic limit, as expected for delocalised states.

The disorder strength denoted throughout as $\wc$ is that above which \emph{all} 
states in the band are MBL (i.e.\ $\wc \equiv \wc(\w$$ =$$0)$,  as band centre states 
are the last to localise). For $W<\wc$, some states in the band will be delocalised, and others MBL -- the spectrum hosts mobility edges. For any such $W$  then, from the above,
delocalised states contribute a factor of $1/2$ to the summand in Eq.\ \ref{eq:rbarinftsum}
as $L\to \infty$, while MBL states contribute a factor strictly $<1/2$.
It is therefore only if all states in the band are delocalised -- or in practice all
but a tiny fraction  -- that  the long-time limit $\overline{r}(\infty)/L$ will be $1/2$.
From Fig.\ \ref{fig:rbar(n)W=1,2}, this indeed appears to be the case for $W=1$. On further increasing $W$, however, a non-negligible fraction of MBL states must arise, resulting in
$\overline{r}(\infty)/L<1/2$. The $W=2$ case in Fig.\ \ref{fig:rbar(n)W=1,2} appears 
to provide an example of this (at least up to the largest $L$ considered here). And
the trend certainly becomes more pronounced with increasing $W$, e.g.\ for $W=3$, 
$\overline{r}(\infty)/L$ is $\lesssim 0.4$  for the largest $L$ studied.

 Eq.\ \ref{eq:rbarinftsum} shows that $\overline{r}(t=\infty)/L$ can be resolved as
a sum over contributions from all eigenstates in the band. This in fact is true for any $t$.
As elaborated in Appx.\ \ref{section:eigresolution}, it arises because $P_{IJ}(t)$ 
can be eigenstate resolved in the form $P_{IJ}(t) =\nh^{-1}\sum_{n} \PIJn{IJ}(t)$, 
with $\PIJn{IJ}(t)$ pertaining to a particular state $n$ of energy $E_{n}$, and
given by
\begin{equation}
\label{eq:PIJntcos}
\nh^{-1}\PIJn{IJ}(t) =\sum_{m}\mathrm{cos}\big[(E_{n}^{\pd}-E_{m}^{\pd})t\big]
A_{nI}^{\pd}A_{nJ}^{\pd}A_{mI}^{\pd}A_{mJ}^{\pd};
\end{equation}
such that for times $\Gamma t \gg 1$, $\PIJn{IJ}(t)$ is controlled by states $m$ lying in a progressively narrowing window $|E_{n}-E_{m}|\ll \Gamma$ in the  vicinity of the chosen energy 
$E_{n}$. We remark in passing that $\nh^{-1}\PIJn{IJ}(t)$ can equally be expressed as an eigenstate expectation value of an operator, see Eq.\ \ref{eq:eigres3}.

Since $\overline{r}(t)$ is linear in the $\{P_{IJ}(t)\}$, it too can be eigenstate
resolved, $\overline{r}(t)=\nh^{-1}\sum_{n}\barrn{}(t)$. In particular, from
Eq.\ \ref{eq:rbarinftsum}, $\barrn{}(\infty)/L =1/[1+e^{1/\xi_{F,n}}]$. 
The lower panels in Fig.\ \ref{fig:rbar(n)W=1,2} show the $t$-dependence of
$\barrn{}(t)/L$ for states $n$ in the immediate vicinity of the band centre, 
with $W=1,2$. Since $W<\wc$ here, one expects the long-time limit of $\barrn{}(t)/L$
for band centre states to be $1/2$, which appears consistent with the data.
For the  $W=2$ example, it is also seen from Fig.\ \ref{fig:rbar(n)W=1,2} that the
regime of slower dynamics mentioned above, setting in above $\Gamma t\sim 1$, is evident in both
$\overline{r}(t)/L$ and $\barrn{}(t)/L$; suggesting that this behaviour is associated with delocalised states in the spectrum.

 \begin{figure}
\includegraphics[width=\columnwidth]{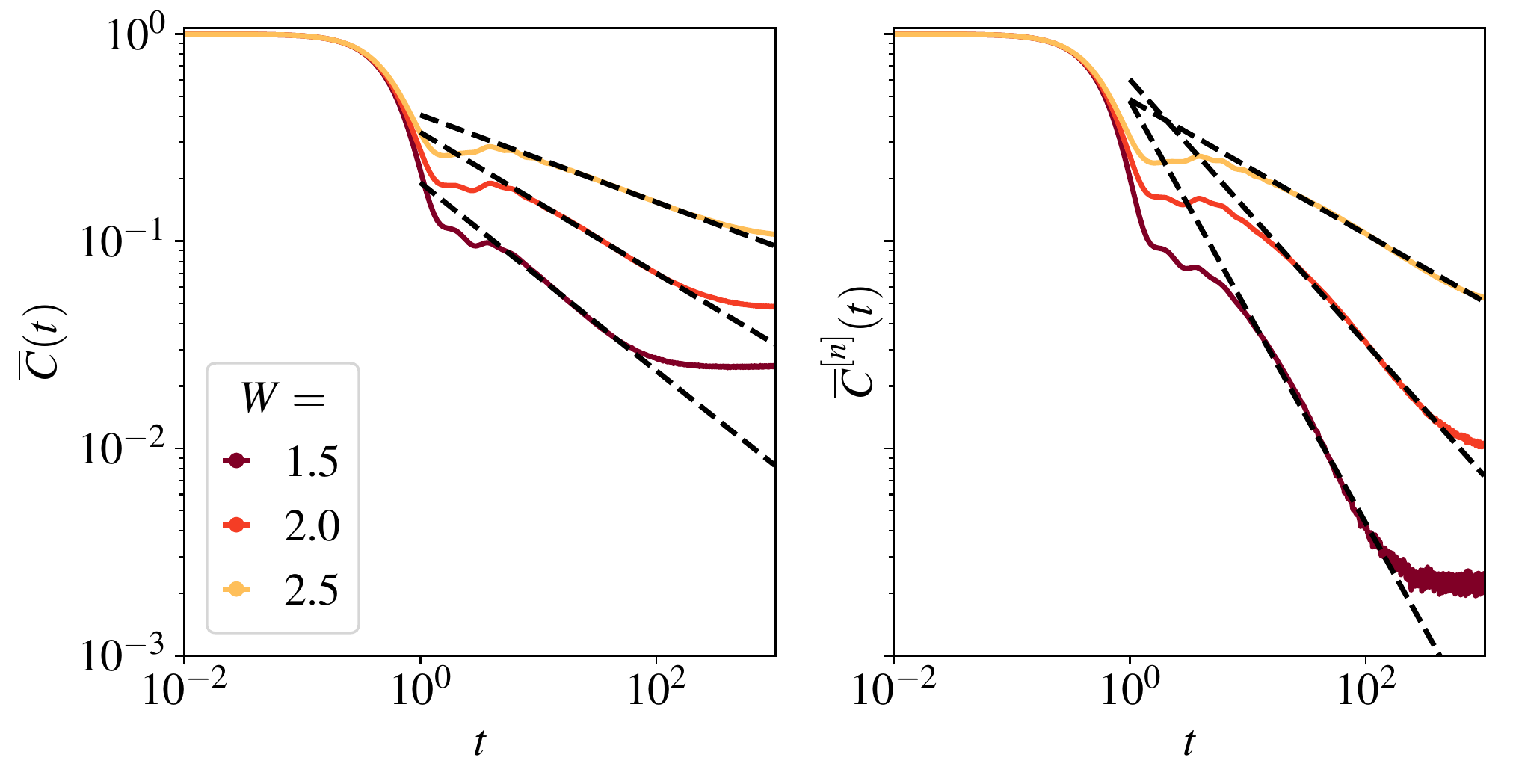}
  \caption{
	For $W=1.5, 2$ and $2.5$, with $L=14$,
	ED results for spin autocorrelation functions
	\emph{vs} $t$, on a log-log scale.
	\emph{Left panel}:  $\overline{\mathcal{C}}(t)=1-2\overline{r}(t)/L$. 
	\emph{Right panel}: $\barCn{}(t)=1-2\barrn{}(t)/L$ for band centre eigenstates $n$.
	In either panel, for each $W$, dashed lines show power-law fits to the intermediate-time behaviour, with the power-law exponents found to decrease with increasing $W$.
	} 
	\label{fig:Cbar-t-ergodic}
\end{figure} 

\begin{figure}
\includegraphics[width=\columnwidth]{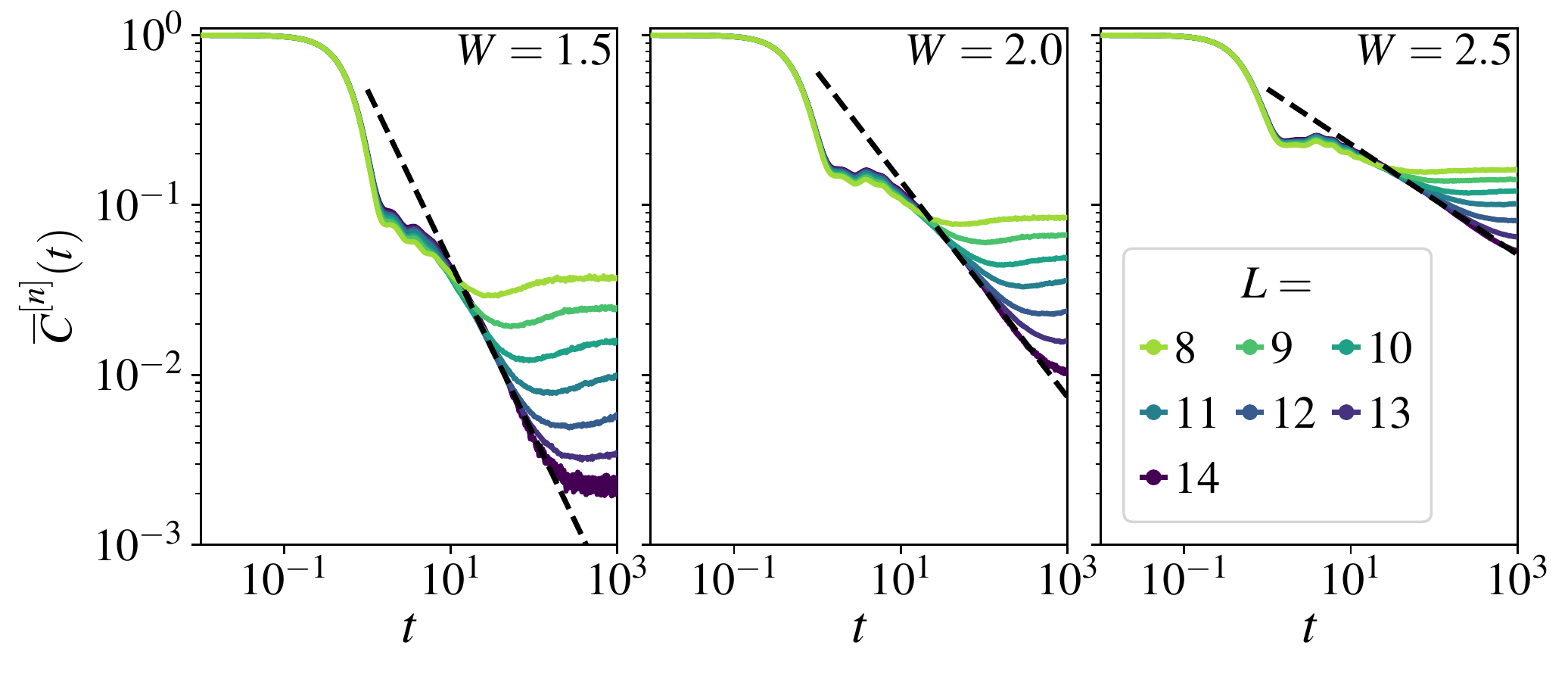}
  \caption{
	For $W=1.5, 2$ and $2.5$, showing  $\barCn{}(t)=1-2\barrn{}(t)/L$ 
	\emph{vs} $t$, for band centre states $n$. Dashed lines show power-law fits to the 
	intermediate-time behaviour, onto which the data scales progressively with increasing $L$.
	} 
	\label{fig:Cbar(n)-t-ergodic}
\end{figure}

To examine further these slow dynamics at intermediate times, we consider equivalently the
$t$-dependence of the spin autocorrelation functions, 
$\overline{\mathcal{C}}(t)=1-2\overline{r}(t)/L$  (Eq.\ \ref{eq:Cfull}) and its 
eigenstate-resolved counterpart $\barCn{}(t)=1-2\barrn{}(t)/L$. The former is shown on a 
log-log scale in the left panel of Fig.\ \ref{fig:Cbar-t-ergodic} for $W=1.5, 2$ and $2.5$, 
with $L=14$.  As seen from the figure, $\overline{\mathcal{C}}(t)$ and hence $\overline{r}(t)/L$ 
exhibits an intermediate-time power-law decay,
$\overline{\mathcal{C}}(t)\propto t^{-\alpha}$ with $\alpha <1$.
With increasing $W$, the exponent $\alpha$ is found to decrease steadily and, subject to the 
usual caveat of modest system sizes, appears to vanish in the vicinity of $W\simeq \wc \sim 4$.
For eigenstates $n$ in the vicinity of the band centre, which are themselves ergodic for $W<\wc$,
the corresponding behaviour of $\barCn{}(t)$ is shown in the right panel of 
Fig.\ \ref{fig:Cbar-t-ergodic}. It too shows an intermediate-time power-law decay, $\barCn{}(t)\propto t^{-\alpha^{\prime}}$ with an exponent $\alpha^{\prime}$ which, while larger than the corresponding $\alpha$ at the same $W$, likewise decreases steadily with increasing $W$ and vanishes around  $\wc$. The $L$-dependence of $\barCn{}(t)$ \emph{vs} $t$ for band centre states
is shown in Fig.\ \ref{fig:Cbar(n)-t-ergodic}, from which the data is seen to scale progressively
further onto the power-law decay with increasing $L$.

Subdiffusive dynamics of the spin autocorrelation function, in the ergodic phase for a wide range of disorder strength preceding the MBL regime, has been extensively studied in models with conserved total magnetisation, such as the disordered XXZ chain~\cite{agarwal2015anomalous,SantosPRB2015,luitz2016long,luitz2016anomalous,luitz2017ergodic,roy2018anomalous}.
In the Ising spin chain \eqref{eq:ham}  considered here, total magnetisation is not by contrast
conserved, so the appearance of subdiffusive dynamics warrants explanation. Indeed, for a Floquet version of the Ising chain \eqref{eq:ham}, a previous numerical study raised the possibility
that the spin autocorrelation decays as a stretched exponential in 
time~\cite{lezama2019apparent}. However, the key point here is that although total magnetisation is not conserved in our model, total energy is (trivially, the Hamiltonian being time 
independent). As a result, the autocorrelator of the local energy density shows subdiffusive dynamics; $\braket{\hat{H}_\ell(t)\hat{H}_\ell}\sim t^{-\alpha^{\prime\prime}}$ where $\hat{H}_\ell\equiv h_\ell\hat{\sigma}^z_\ell + \Gamma\hat{\sigma}^x_\ell+\tfrac{1}{2} [J_\ell\hat{\sigma}^z_\ell\hat{\sigma}^z_{\ell+1} + J_{\ell-1}\hat{\sigma}^z_{\ell-1}\hat{\sigma}^z_{\ell}]$. 
The spin operator $\hat{\sigma}^z_\ell$ is not however  orthogonal to the local energy density operator, $\mathrm{Tr}[\hat{\sigma}^z_\ell \hat{H}_\ell]\neq 0$. Therefore at intermediate to late times, the spin autocorrelation picks up the (sub)diffusive tails emerging from the autocorrrelator of the local energy density; explaining physically the origin of the power-law decay of the spin autocorrelator.


\subsection{$\boldsymbol{\overline{r}(t)}$: MBL regime}
\label{subsection:rbar(t)mbl}

 \begin{figure}
\includegraphics[width=\columnwidth]{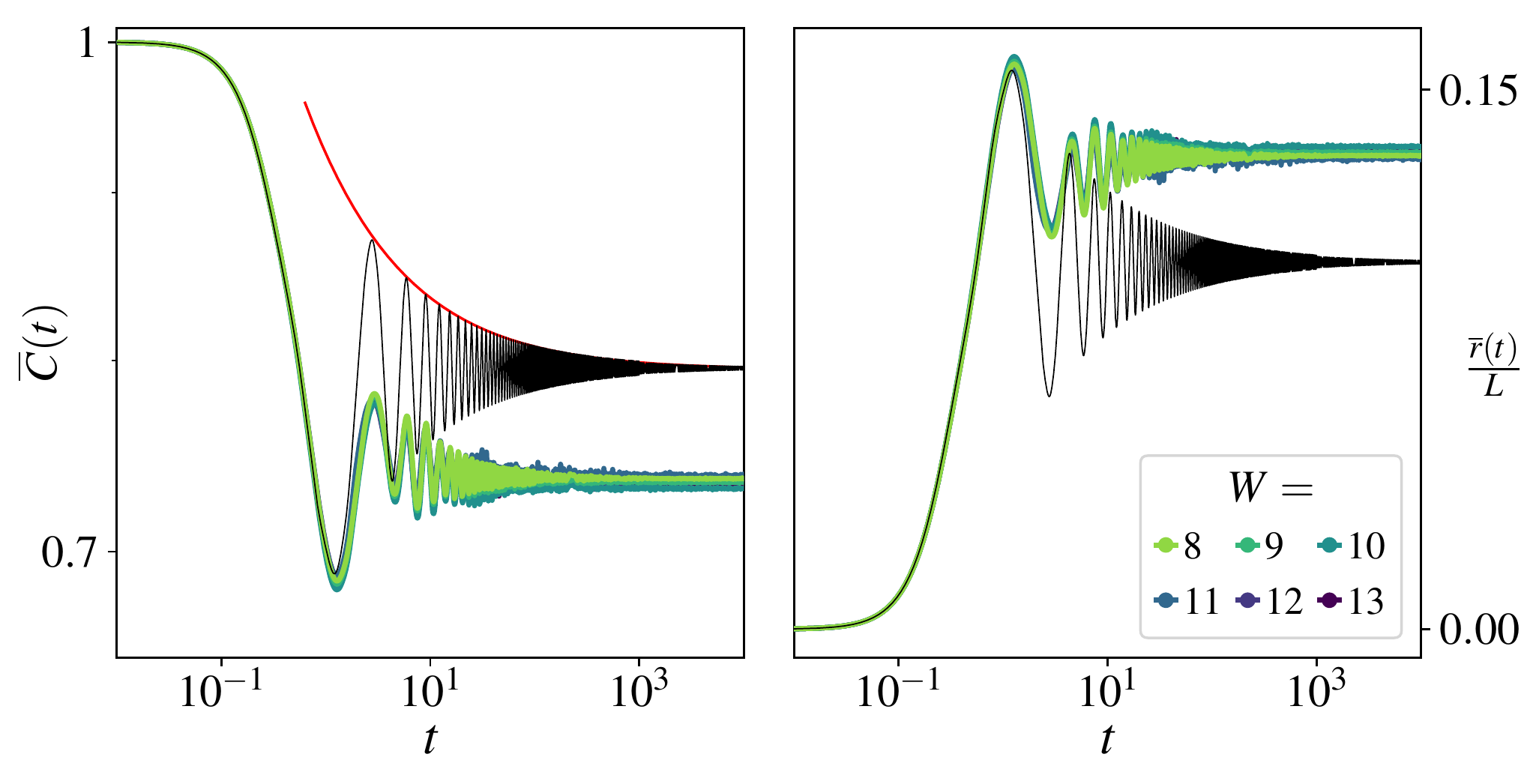}
  \caption{
	For $W=7$, 	ED results for the spin autocorrelation function 
	$\overline{\mathcal{C}}(t)=1-2\overline{r}(t)/L$ (left panel) and $\overline{r}(t)/L$ 
	itself (right panel), \emph{vs} $t$ ($\Gamma \equiv 1$) and for
	the system sizes $L$ indicated. Solid black line shows for comparison the
	corresponding exact result for $\mblo$ Eqs.\ \ref{eq:rbarandcbarmblo},\ref{eq:z0def}; 
	red line in the left panel gives the asymptotic behaviour Eq.\ \ref{eq:cbarmbloasym}. 
	} 
	\label{fig:Cbar-t-mbl}
\end{figure}

To illustrate results in  the MBL regime, Fig.\ \ref{fig:Cbar-t-mbl} shows the spin autocorrelation function $\overline{\mathcal{C}}(t)$, and  $\overline{r}(t)/L$ itself, 
for  disorder strength $W=7$. The behaviour seen is representative of the MBL regime for 
$W\gtrsim 4.5$ or so, and qualitatively different from that characteristic of the ergodic regime.

$\overline{\mathcal{C}}(t)=1-2\overline{r}(t)/L$ in Fig.\ \ref{fig:Cbar-t-mbl}
shows clear damped oscillatory behaviour.  It plateaus to a non-zero long-time value 
($\sim 0.7$, well above zero), indicative of persistent memory of initial conditions; and is barely $L$-dependent over the range studied.
Equivalently, the long-time limit of $\overline{r}(t)/L$ is $\ll 1/2$ (as seen also in 
Fig.\ \ref{fig:Pbarrtmap} for the mode of $\overline{P}(r;t)$ at long times).
This in turn is consistent with Eq.\ \ref{eq:rbarinftsum} above, where, 
with all states $n$ MBL for $W>\wc$, all correlation lengths $\xi_{F,n}$ are finite and 
hence $\overline{r}(\infty)/L <1/2$.

Two further points about Fig.\ \ref{fig:Cbar-t-mbl} should be made at this stage, each of which merits some understanding (Sec.\ \ref {subsection:mblo1} below). First, while the long-time behaviour is seen to be reached in practice by $\Gamma t\sim 10^{2}$, damped oscillations about that limit set in at shorter times $\Gamma t \sim \mathcal{O}(1)$, above which 
the envelope of the oscillation is in fact rather well fit by a power-law decay
$\propto t^{-\beta}$ with $\beta \approx 1/2$. Second, in parallel to 
Sec.\ \ref{subsection:rbar(t)} for the ergodic phase, in the MBL regime one can equally consider the eigenstate-resolved  $\barCn{}(t)=1-2\barrn{}(t)/L$, e.g.\ for states $n$ in the vicinity of the band centre. On doing that, one finds essentially no discernible difference from the results for $\overline{\mathcal{C}}(t)$ shown in Fig.\ \ref{fig:Cbar-t-mbl}.


\subsubsection{$\boldsymbol{\mblo}$}
\label{subsection:mblo1}

To obtain an understanding of the above results, we consider now what we refer to as
`$\mblo$'~\cite{roy2021fockspace}. Sufficiently deep in the MBL phase, the model 
\eqref{eq:ham} is perturbatively connected to the non-interacting limit 
$J_{\ell}=0$ ($\mblo$). Here, although the system is `trivially' MBL -- because 
$\mathcal{H}$  (Eq.\ \ref{eq:ham}) is site-separable in real-space and the system a 
set of noninteracting spins -- the behaviour on the Fock space is 
known~\cite{roy2021fockspace} to be non-trivial, and the Fock-space $\mathcal{H}$ 
Eq.\ \ref{eq:hamTBM} remains fully connected on the graph.

As outlined in Appx.\ \ref{section:mblo}, for $\mblo$ the exact disorder-averaged 
$\overline{P}_{IJ}(t)$ can be obtained, starting from the basic definition 
$P_{IJ}(t)=|G_{IJ}(t)|^{2}$, Eq.\ \ref{eq:PIJdef}.
With $J,I$ any pair of FS sites separated by a Hamming distance $r_{IJ}=r$,
the result is 
\begin{equation}
\label{eq:PIJmblo}
\overline{P}_{IJ}^{\pd}(t) ~=~[z_{0}^{\pd}(t)]^{r}[1-z_{0}^{\pd}(t)]^{(L-r)}
\end{equation}
with $z_{0}(t)$ given by
\begin{equation}
\label{eq:z0def}
\begin{split}
z_{0}^{\pd}(t)~=&~\Big\langle \frac{\Gamma^{2}}{h^{2}+\Gamma^{2}}\mathrm{sin}^{2}
\big(\sqrt{h^{2}+\Gamma^{2}}~t\big)
\Big\rangle_{\dis}^{\pd}
\\
=&~\int_{0}^{W}\frac{dh}{W}~\frac{\Gamma^{2}}{h^{2}+\Gamma^{2}}\mathrm{sin}^{2}
\big(\sqrt{h^{2}+\Gamma^{2}}~t\big).
\end{split}
\end{equation}
Note that $\overline{P}_{IJ}(t)$ again has the binomial form Eq.\ \ref{eq:PIJrbinom},
deduced on general grounds in Sec.\ \ref{section:shortt} for short-times;
and Eq.\ \ref{eq:z0def} obviously recovers, as it ought, the known asymptotic behaviour
$z_{0}(t) = (\Gamma t)^{2}$ as $\Gamma t\to 0$.

Eq.\ \ref{eq:PIJmblo} in fact depends solely on the Hamming distance $r$ and is otherwise
independent of the particular FS sites $J,I$ (see  Appx.\ \ref{section:mblo}).
Hence, from Eq.\ \ref{eq:Prtdef}, $\overline{P}_{I}(r;t)$ is independent of $I$, 
and $\overline{P}(r;t)\equiv \overline{P}_{I}(r;t)$ is thus
\begin{equation}
\label{eq:Pbarmblo}
\overline{P}(r;t)
~=~\binom{L}{r}[z_{0}^{\pd}(t)]^{r}[1-z_{0}^{\pd}(t)]^{(L-r)}.
\end{equation}
From this follows the first moment $\overline{r}(t)$ and hence $\overline{\mathcal{C}}(t)$,
\begin{equation}
\label{eq:rbarandcbarmblo}
\frac{\overline{r}(t)}{L}=~z_{0}^{\pd}(t), ~~~~
\overline{\mathcal{C}}(t) =1-2z_{0}^{\pd}(t),
\end{equation}
each of which is $L$-independent for all $t$. Fig.\ \ref{fig:Cbar-t-mbl} compares this result 
for $\overline{\mathcal{C}}(t)$ and $\overline{r}(t)/L$, to ED results for the interacting case. The strong qualitative parallels between the two are self evident.

$\mblo$ is fully determined by $z_{0}(t)$ (Eq.\ \ref{eq:z0def}) which, via
the double-angle formula for $\mathrm{sin}^{2}\theta$ (and setting $\Gamma \equiv 1$) is
\begin{equation}
\label{eq:z0itop}
z_{0}^{\pd}(t)~=~p -\tfrac{1}{2}K(t), ~~~~
p~=~\frac{\mathrm{tan}^{-1}(W)}{2W}
\end{equation}
 with
$K(t)=\langle (h^{2}+1)^{-1}\mathrm{cos}\big(2\sqrt{h^{2}+1}~t \big)\rangle_{\dis}$.
$K(t)$ vanishes as $t\to \infty$ (see Eq.\ \ref{eq:K(t)asymp}).  The long-time limits are then
$\overline{r}(\infty)/L =p$ and $\overline{\mathcal{C}}(\infty)=1-2p$;
with $p<1/2$ necessarily such that $\overline{\mathcal{C}}(\infty)>0$ and 
$\overline{r}(\infty)/L<1/2$, as characteristic of an MBL phase.
For $W=7$, as in Fig.\ \ref{fig:Cbar-t-mbl}, the $\mblo$ 
$\overline{\mathcal{C}}(\infty)\simeq 0.8$, only slightly larger than its
interacting counterpart of $\overline{\mathcal{C}}(\infty)\simeq 0.7$.

We add that in Appx.\ \ref{section:mblo} we also point out the connection  
$\overline{P}(r;\infty)\equiv \overline{F}_{n}(r)$
between the long-time limit of $\overline{P}(r;t)$
and the eigenstate correlation function $\overline{F}_{n}(r)$
(Eq.\ \ref{eq:Fnbardef}, which for $\mblo$ is the same for all eigenstates $n$).

The $t$-dependence of $K(t)$ is readily determined. 
Its asymptotic behaviour, formally for $t\gg 1$, is given by
\begin{equation}
\label{eq:K(t)asymp}
K(t)~\sim ~\frac{1}{2W}\sqrt{\frac{\pi}{2}}~
\frac{[\mathrm{cos}(2t)-\mathrm{sin}(2t)]}{\sqrt{t}},
\end{equation}
vanishing as a power-law $\propto 1/\sqrt{t}$ superimposed on the oscillating envelope 
of period $\pi$. The maxima of the oscillatory part occur at the discrete set of points
$t=\tfrac{7}{8}\pi +\pi n $ ($n\in \mathbbm{N}_{0}$), at which 
\begin{equation}
\label{eq:cbarmbloasym}
\overline{\mathcal{C}}(t) \sim \left(1 -\frac{\mathrm{tan}^{-1}(W)}{W}\right)
+\frac{\sqrt{\pi}}{2W}\frac{1}{\sqrt{t}}
\end{equation}
This is superimposed on the $\mblo$ result for $\overline{\mathcal{C}}(t)$ shown
in Fig.\ \ref{fig:Cbar-t-mbl}, and in practice is seen to account very well for the 
behaviour down to times $t$ on the order of unity.

For $\mblo$ one can also determine the eigenstate-resolved
$\barCn{}(t)=1-2\barrn{}(t)/L$ for an arbitrary eigenstate $|n\rangle$.
In this case, reflecting the real-space site-separability of $\mathcal{H}$, it can be shown
(although we do not prove it here) that the disorder-averaged
$\barClln{}(t) =\overline{\langle n|\hat{\sigma}_{\ell}^{z}(t)\hat{\sigma}_{\ell}^{z}|n\rangle}$
is independent of both the site $\ell$ and the particular eigenstate $|n\rangle$.
In consequence, $\barCn{}(t)\equiv \overline{\mathcal{C}}(t)$;
providing a rationale for the fact,  mentioned in Sec.\ \ref{subsection:rbar(t)mbl} above, 
that our ED calculations of $\barCn{}(t)$ in the interacting case are barely  
discernible from those for $\overline{\mathcal{C}}(t)$.

\subsubsection{$\mathbf{Fluctuations}$}
\label{subsection:flucs}

While our primary focus has been the first moment of $\overline{P}(r;t)$,
higher central moments are also of course calculable. As previously mentioned, 
the fact that it is $\overline{r}(t)/L$ which generically remains finite in the 
thermodynamic limit means that it is fluctuations in this quantity one should consider,
as reflected in $\sigma^{2}(t):=\overline{\delta r^{2}}(t)/L^{2}$
(with  $\overline{\delta r^{2}}(t)$ from Eq.\ \ref{eq:dealtar2def}).
As shown in Sec.\ \ref{section:shortt} for the short-$t$ domain, which holds for all
disorder/interaction strengths, $\sigma^{2}(t)\propto 1/L$.
Fluctuations are thus entirely suppressed in the thermodynamic limit.
Just the same situation arises for $\mblo$ which, from Eq.\ \ref{eq:Pbarmblo}
for $\overline{P}(r;t)$, gives $\sigma^{2}(t)=z_{0}(t)[1-z_{0}(t)]/L$.
Indeed, employing steepest descents on Eq.\ \ref{eq:Pbarmblo} shows $\overline{P}(r;t)$ as a function of $y\equiv r/L$ to be Gaussian, with a mean of $z_{0}(t)$ ($=\overline{r}(t)/L$) and 
variance $\sigma^{2}(t)\propto 1/L$; such that it becomes $\delta$-distributed in the
thermodynamic limit.

More generally, across essentially the full range of disorder strengths, our ED calculations 
are also qualitatively consistent with the above conclusions: aside from a small $W$-interval around $\wc \sim 4$, with increasing $L$ we find $\overline{\delta r^{2}}(t)/L^{2}$ to progressively decrease, and the $\overline{P}(r;t)$ profile to narrow.

\section{Lateral probability transport}
\label{section:lateraltransportmain}

We turn now to the substantive question of how the $t$-dependent wavefunction spreads out laterally,  as reflected in the time-dependent distribution of probabilities across the rows of the FS graph.

 \begin{figure}
\includegraphics[width=\columnwidth]{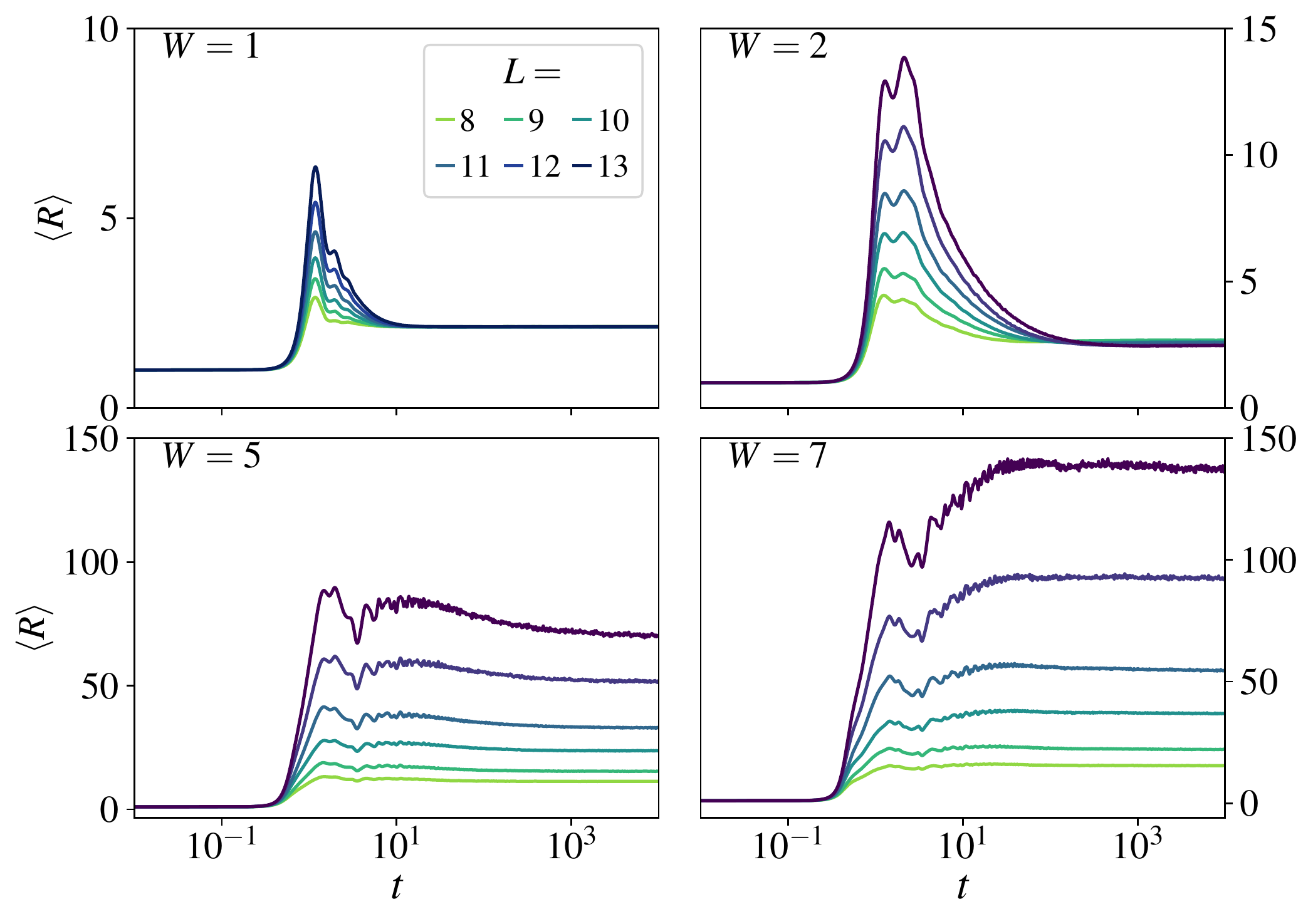}
  \caption{ $t$-dependence of the average $\langle R\rangle$ 
	($=N_{r}\langle\mathcal{I}_{2}\rangle$), see Eqs.\ \ref{eq:RIdef},\ref{eq:Rmeandef},\ref{eq:IPRandR}.
	Shown  for $r=L/2$ with $W=1,2$ (top row) and
	$W=5,7$ (bottom row), for system sizes indicated.
	} 
	\label{fig:Average_R_t}
\end{figure}

As explained in Sec.\ \ref{subsection:lateraltransportA}, the quantity 
$R_{I}(r;t)$ (Eq.\ \ref{eq:RIdef}) provides a natural measure of fluctuations in the 
distribution of $P_{IJ}$'s along any given row $r$, and is related directly to  the 
row-resolved, $t$-dependent IPR by $R_{I}(r;t)=N_{r}\mathcal{I}_{I,2}(r;t)$ 
(Eq.\ \ref{eq:IPRandR}), with $N_{r}=\binom{L}{r}$ the number of FS sites on row $r$.
We consider first the averages, $\langle R\rangle \equiv \langle R\rangle(r;t)$ or 
$\langle \mathcal{I}_{2}\rangle$  (Eqs.\ \ref{eq:Rmeandef},\ref{eq:IPRandR}), over disorder realisations and FS sites $I$, before turning to the full probability distribution 
$P_{R}(x)$ (Eq.\ \ref{eq:PdistR}) of $R_{I}(r;t)$.

There is no \emph{a priori} requirement here to average over all FS sites $I$, so
in this section we choose (for numerical convenience) to average over FS sites $I$ 
whose site energies $\mathcal{E}_{I}$ lie close to their mean value of zero.
Since our interest lies in dynamics, we also focus on a particular, representative 
$r$ throughout the section. We choose the midpoint of the FS graph, $r=L/2$ (or $r=(L-1)/2$ for odd $L$), and have checked that the key results arising are not dependent on this choice.
Fig.\ \ref{fig:Average_R_t} shows the $t$-dependence of $\langle R\rangle$ for the
system sizes indicated, with $W=1,2$ representative of the ergodic regime and $W=5,7$ 
of the MBL regime.

Three notable points are evident in Fig.\ \ref{fig:Average_R_t}. First, in the short-time domain 
$t\lesssim 0.1$, $\langle R\rangle =1$ independently of $L$, for \emph{all} interaction strengths 
$W$. As pointed out in Sec.\ \ref{subsection:lateraltransportA} (under Eq.\ \ref{eq:RIdef}), 
this is the limit of complete homegeneity, where all $P_{IJ}(t)$'s on any given row
of the graph are the same (itself shown in Sec.\ \ref{section:shortt}).
In consequence, $R_{I}(r;t)=1$ (Eq.\ \ref{eq:RIdef}) and hence $\langle R\rangle =1$, as seen;
equivalently the row-resolved IPR 
$\langle \mathcal{I}_{2}\rangle =N_{r}^{-1}\langle R\rangle = N_{r}^{-1}$.

Second, consider now the opposite limit in Fig.\ \ref{fig:Average_R_t}, viz.\ the long-time behaviour. For $W=1,2$, $\langle R\rangle$ here is $\mathcal{O}(1)$ and $L$-independent, just as it is in the short-time domain; while for $W=5,7$ by contrast, 
$\langle R\rangle$ clearly grows with increasing $L$. As explained in 
Sec.\ \ref{subsection:lateraltransportA} (under Eq.\ \ref{eq:IPRandR}),
the former behaviour again reflects the essentially uniform distribution of probabilities 
$P_{IJ}(t)$ over FS sites on the row, with
$\langle \mathcal{I}_{2}\rangle =N_{r}^{-1}\langle R\rangle \propto N_{r}^{-1}$,
as one expects for an ergodic regime at late times. In the MBL regime by contrast, the growth 
of $\langle R\rangle$ with  $L$ reflects that probabilities, and hence the wavefunction, 
are strongly inhomogeneously distributed on the row.

\begin{figure}
\includegraphics[width=\columnwidth]{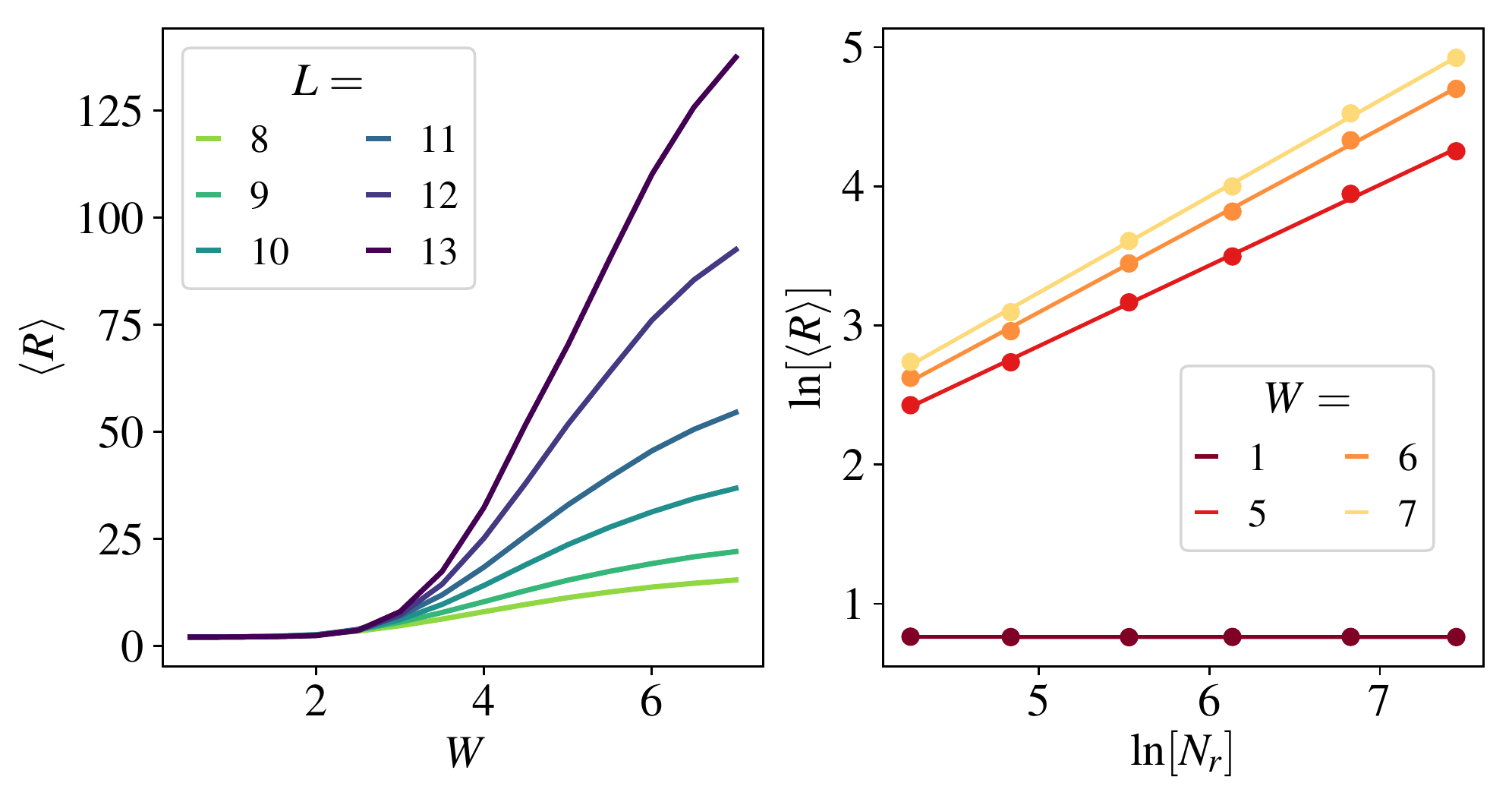}
  \caption{Results here refer to long-time behaviour (taken at $t=10^{4}$). 
	\emph{Right panel}: $\ln\langle R\rangle$ \emph{vs} $\ln N_{r}$, shown for
	$W=5,6,7$ and $W=1$;
showing the scaling behaviour $\langle R\rangle \sim N_{r}^{1-\nu}$
	$\equiv$ $\langle\mathcal{I}_{2}\rangle \sim N_{r}^{-\nu}$, with exponent $\nu <1$ in 
	the MBL regime and $\nu =1$ in the ergodic regime.
	\emph{Left panel}: $\ln\langle R\rangle$ \emph{vs} $W$ for system sizes indicated.
	} 
	\label{fig:R_average_Longt_W}
\end{figure}

As was conjectured on physical grounds in Sec.\ \ref{subsection:lateraltransportA},
the $L$-dependence of $\langle R\rangle$ in the MBL regime is indeed
found to be $\langle R\rangle\sim N_{r}^{1-\nu}$ -- or equivalently
$\langle \mathcal{I}_{2}\rangle \sim N_{r}^{-\nu}$ for the row-resolved IPR -- with a
long-time (multi)fractal exponent $\nu <1$.  That this is so is demonstrated in 
Fig.\ \ref{fig:R_average_Longt_W}, right panel, where in the MBL regime the long-time exponent
$\nu$ is also seen to decrease with increasing $W$ (for $W=5,7$, $\nu \simeq 0.4, 0.3$ 
respectively). This figure also shows the same plot for $W=1$, confirming the $L$-independence 
of $\langle R\rangle$ (corresponding to $\nu =1$). The left panel of 
Fig.\ \ref{fig:R_average_Longt_W} gives the late-time $\langle R\rangle$ as a function of $W$, 
confirming both the strong $L$-dependence inside the MBL regime, and its corresponding
absence in the ergodic phase. Equally, it shows a typical `crossover $W$-window',
whose presence is inevitable given accessible system sizes; and which, without 
further detailed scaling analysis, precludes substantive consideration of $W$'s in the vicinity of $\wc \sim 3.8$~[\onlinecite{abanin2021distinguishing}] (which is not our aim here).

Third, consider again Fig.\ \ref{fig:Average_R_t} for the ergodic phase $W$'s.
Although as above $\langle R\rangle \equiv \langle R\rangle (t)$ is
$L$-independent at both short- and long-times, for times $t$ on the order of unity 
$\langle R\rangle (t\simeq 1)$ shows a strong  $L$-dependence. This too is found to have 
the form $\langle R\rangle (t\simeq 1)\sim N_{r}^{1-\nu}$, directly analogous to 
Fig.\ \ref{fig:R_average_Longt_W} (right panel), and with an exponent
$\nu \equiv \nu(t\simeq 1)$ that likewise decreases with increasing $W$ (for $W=1,2$, $\nu(t=1)\simeq 0.8$ and $0.7$ respectively).

The overall physical picture arising from the above is then as follows. Following the 
$W$-independent, short-time complete homogeneity of the squared wavefunction 
amplitudes/probabilities along the row of the graph,  the $L$-dependence arising  by $t\sim 1$ -- again for all $W$ -- indicates the dynamical emergence of multifractal behaviour of the wavefunction. The latter persists with increasing $t$ in the MBL regime, until by $t \sim 10$ or so the long-time multifractality is well established. For the ergodic $W$'s by contrast, that evolution is arrested; and the system instead crosses over from incipient multifractality to the ergodic behaviour reflected in $\langle R \rangle\sim N_{r}^{0}\sim \mathcal{O}(1)$
(i.e.\ $\langle \mathcal{I}_{2}\rangle \sim N_{r}^{-1}$), indicating an essentially uniform distribution of probabilities along the row. This picture, pertaining to the row-resolved IPR,
provides a rather natural and consistent complement to that shown in
Sec.\ \ref{subsection:shorttmultif} to arise for the behaviour of the
conventional IPR over the full Fock-space (see Fig.\ \ref{fig:tau2t-dep}).


\subsection{Probability distributions}
\label{subsection:vertprobdists}

\begin{figure*}
\includegraphics[width=0.9\linewidth]{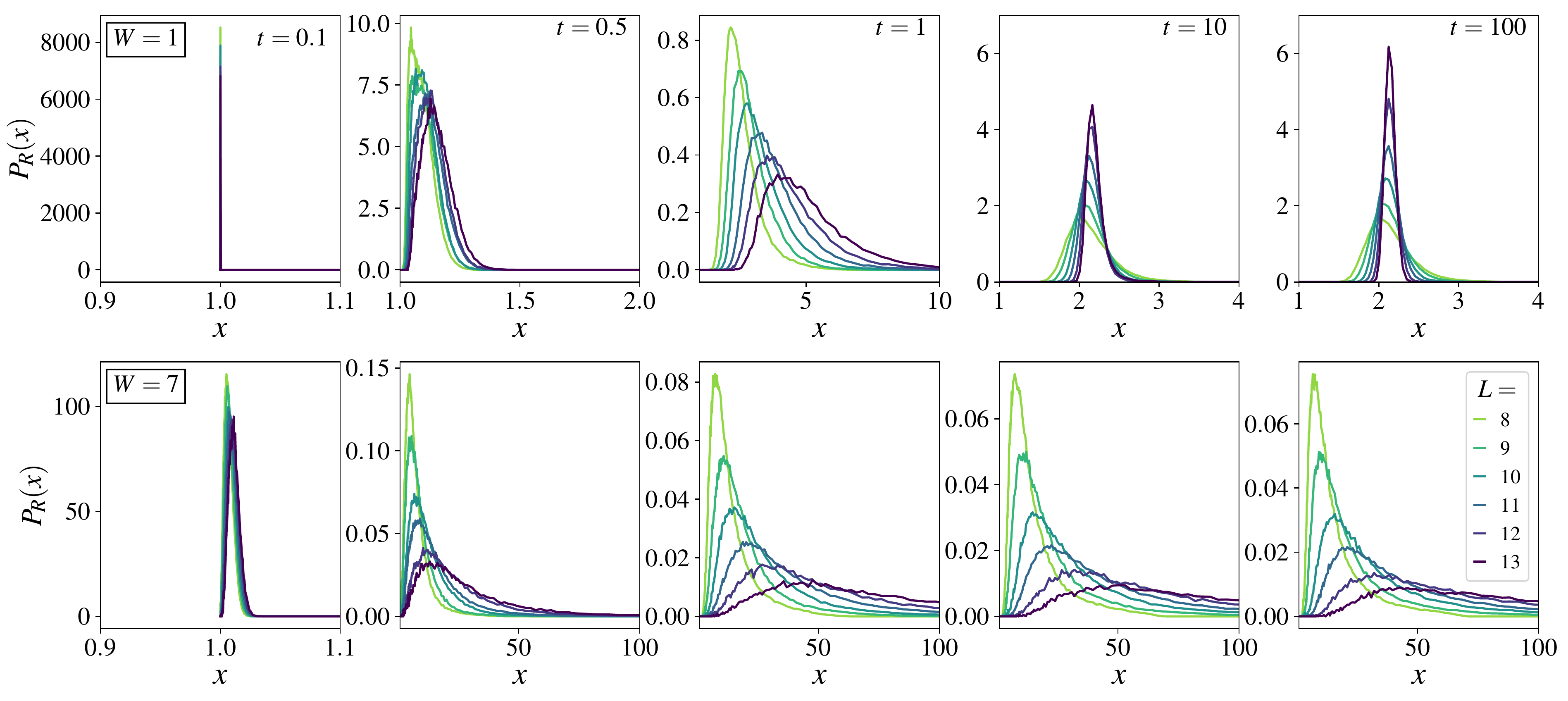}
\caption{Probability distribution $P_{R}(x)$ of $x$ $(\equiv R_{I}(r;t))$
for $W=1$ (top row) and $W=7$ (bottom), at the sequence of $t$'s specified,
and for different system sizes $L$ as indicated. Full discussion in text.
}
\label{fig:P_R}
\end{figure*}

The discussion above has centred on the average value, $\langle R\rangle$, of
$R_{I}(r;t) =N_{r}\mathcal{I}_{I,2}(r;t)$ (Eq.\ \ref{eq:RIdef}). Now we 
consider the full probability distribution of $R_{I}(r;t)$, given by
(Eq.\ \ref{eq:PdistR}) $P_{R}(x)=\langle\delta (x-R_{I}(r;t))\rangle_{\dis,I}$
(with the $I$-averaging over sites whose FS site-energies are close to their mean, as
mentioned above); and the first moment of which distribution is precisely
$\langle R\rangle$ considered above.

To illustrate the key points here, Fig.\ \ref{fig:P_R} shows $P_{R}(x)$ \emph{vs} $x$
for $W=1$ (top row) and $W=7$ (bottom), at the sequence of $t$'s indicated, and over the 
range of system sizes studied. Note that, for either $W$, $P_{R}(x)$ for $t=0.1$ 
is $\delta$-distributed at $x=1$. This reflects the short-time regime ($t\lesssim 0.1$) for which, as discussed above in relation to Fig.\ \ref{fig:Average_R_t}, $R_{I}(r;t)=1$ (for any
$r,I$ and all $W$), and hence $P_{R}(x)=\delta (x-1)$.

First consider $W=1$ in Fig.\ \ref{fig:P_R}, illustrating the ergodic regime. For any given 
$L$,  the 
$P_{R}(x)$ distribution evolves most significantly with $t$ over the interval 
$0.1 \lesssim t \lesssim 1$. On further increasing $t$, the mean ($=\langle R\rangle$)
of $P_{R}(x)$ decreases, as also evident from Fig.\ \ref{fig:Average_R_t}.
By $t=100$ the mean of the evidently symmetrical $P_{R}(x)$ appears rather well converged 
over the accessible $L$-range; and the distribution is both narrow and sharpening  with 
increasing $L$ (indeed that behaviour is evident by $t\sim 10$). A simple fit to $P_{R}(x)$ 
for $t=100$, shows it clearly to be normally distributed, with a variance decreasing
with $L$.

The situation is quite different in the MBL regime, illustrated by $W=7$ in Fig.\ \ref{fig:P_R}. Here again, $P_{R}(x)$ evolves most significantly with $t$ over the interval 
$0.1 \lesssim t \lesssim 1$.  For fixed $L$, the distributions are in fact practically 
converged to their long-time limit by $t\sim 1$, above which little  further temporal evolution occurs. Clearly, however, the late-time $P_{R}(x)$ is much broader than its counterpart in the ergodic regime (note the the greatly increased $x$-scale compared to $W=1$),
reflecting the substantial inhomogeneity arising in the MBL regime, as discussed above.
With increasing $L$ the mean and mode of $P_{R}(x)$ continue to increase, as 
discussed in regard to Fig.\ \ref{fig:Average_R_t}. And the distribution is not only visibly 
broad but appears to be heavy-tailed.

 \begin{figure}
\includegraphics[width=\columnwidth]{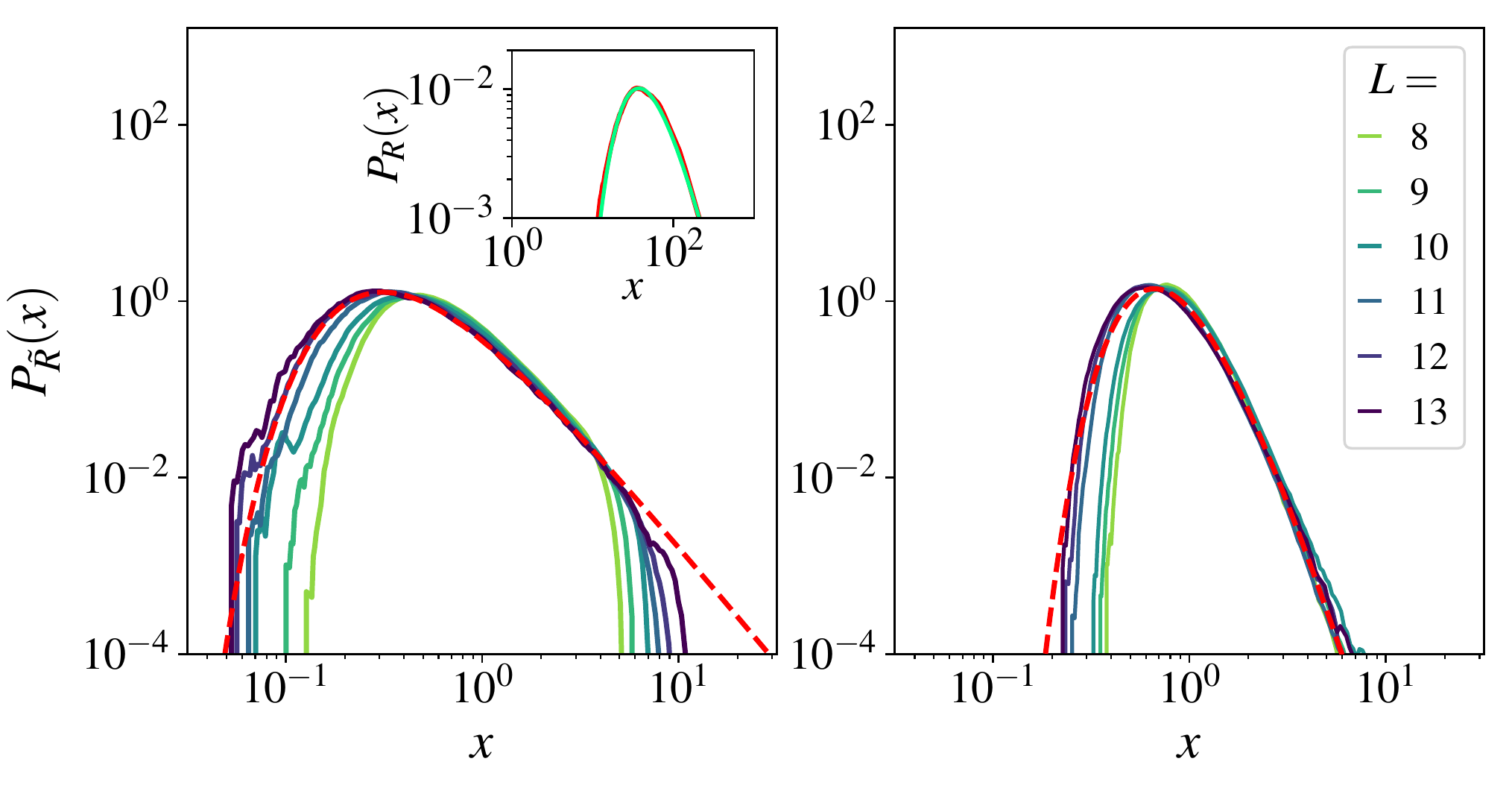}
  \caption{ 
	\emph{Left panel}: For $W=7$ at $t=100$, showing $P_{\tilde{R}}(x)$ (Eq.\ \ref{eq:PRtildedef})
	\emph{vs} $x$ for $L$-values indicated. Dashed line shows comparison to the corresponding
	L\'evy distribution $P_{\tilde{R},\mathrm{L\acute{e}vy}}(x)$ (Eq.\ \ref{eq:PRtildeLevy}).
	Inset: $P_{R}(x)$ for $L=13$, compared to a two-parameter fit 
	 to $P_{R,\mathrm{L\acute{e}vy}}(x)$ (Eq.\ \ref{eq:PRLevy}).
	\emph{Right panel}: Now for $W=2$ at $t=1$, showing $P_{\tilde{R}}(x)$ \emph{vs} $x$.
	Dashed line again compares to corresponding $P_{\tilde{R},\mathrm{L\acute{e}vy}}(x)$.
	} 
	\label{fig:P_R_prime_fit}
\end{figure}

To obtain some understanding of the form of $P_{R}(x)$ in the MBL regime, note first that, in contrast to the ergodic phase, the mean $\langle R\rangle$ of $P_{R}(x)$ is itself increasing with  $L$ (as per Fig.\ \ref{fig:Average_R_t}). To distill this out from the large-$x$ tail of 
$P_{R}(x)$, we thus consider the distribution 
\begin{equation}
\label{eq:PRtildedef}
P_{\tilde{R}}^{\pd}(x)~=~\big\langle\delta\big(x- \tilde{R}\big)\big\rangle_{\dis,I}^{\pd}
~~~~:~\tilde{R}=\frac{R_{I}(r;t)}{\langle R\rangle}
\end{equation}
of  $\tilde{R}=R_{I}(r;t)/\langle R\rangle$, which has a mean of unity for all $t$.
This is shown in Fig.\ \ref{fig:P_R_prime_fit} (left panel) from which, given the modest
accessible $L$-range, reasonable scaling behaviour is seen;  and showing a power-law tail 
$P_{\tilde{R}}(x)\sim x^{-\alpha}$ with $\alpha \simeq 2.5$, such that the variance of the distribution, and all higher moments, are unbounded.


\begin{figure*}
\includegraphics[width=0.9\linewidth]{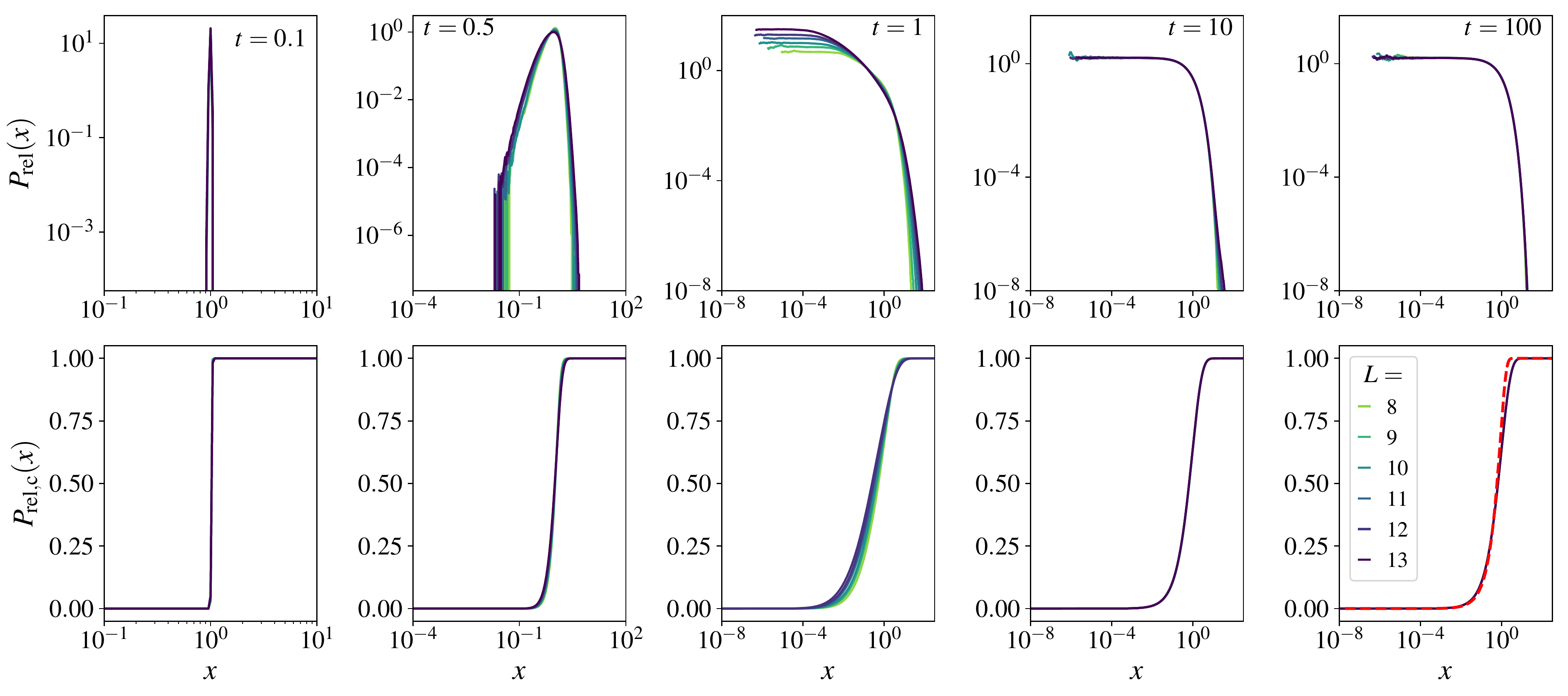}
\caption{For $W=1$. 
Top panels: distribution $\prel(x)$ of $x$ (Eq.\ \ref{eq:defx}) \emph{vs} $x$, at the 
$t$'s specified, and for system sizes $L$ indicated. Bottom panels: cumulative distribution, 
$\prelc(x)$. Red dashed line in final panel shows half-Gaussian fit to data (see text).
}
\label{fig:P_rel_W_1}
\end{figure*}


It appears in fact that $P_{R}(x)$ itself is described by a generalised L\'evy distribution,
\begin{equation}
\label{eq:PRLevy}
P_{R,\mathrm{L\acute{e}vy}}^{\pd}(x)~=~
\frac{A^{\alpha -1}}{\Gamma(\alpha-1)}
\frac{1}{x^{\alpha}}\exp\big(-A/x\big)
~\overset{x\gg A}{\propto}~x^{-\alpha}
\end{equation}
(with $A$ an $L$-dependent constant and $\Gamma(z)$ the gamma function); and  
which heavy-tailed distribution is stable provided $\alpha <3$. The mode of 
$P_{R,\mathrm{L\acute{e}vy}}(x)$ is $x_{\mathrm{mode}}=A/\alpha$ and,
provided $\alpha>2$, its mean is finite and given by 
$\overline{x}=A \Gamma(\alpha-2)/\Gamma(\alpha -1):=A/f(\alpha)$.
The corresponding L\'evy distribution for $\tilde{R}$ is then
\begin{equation}
\label{eq:PRtildeLevy}
P_{\tilde{R},\mathrm{L\acute{e}vy}}^{\pd}(x)~=~
\frac{[f(\alpha)]^{\alpha-1}}{\Gamma(\alpha -1)}~x^{-\alpha}
\exp\big(-f(\alpha)/x\big),
\end{equation}
and depends solely on $\alpha$ and not on $A$. The inset to the left panel of
Fig.\ \ref{fig:P_R_prime_fit}  shows  $P_{R}(x)$ itself (for $L=13$ at $t=100$), 
compared to a two-parameter fit (viz.\ $\alpha, A$) to  $P_{R,\mathrm{L\acute{e}vy}}(x)$
(leading to  $\alpha \simeq 2.5$); the agreement is rather good.
The left panel of Fig.\  \ref{fig:P_R_prime_fit}  shows $P_{\tilde{R}}(x)$
for increasing system sizes, compared to the corresponding $P_{\tilde{R},\mathrm{L\acute{e}vy}}(x)$ Eq.\ \ref{eq:PRtildeLevy}
(dashed line). We add that the $L$-dependence of the fit $P_{R,\mathrm{L\acute{e}vy}}(x)$ itself arises largely from the $L$-dependence of $A$; by contrast, over the accessible $L$-window, 
$\alpha$ varies relatively little (and for which reason 
$P_{\tilde{R},\mathrm{L\acute{e}vy}}(x)$ in Fig.\ \ref{fig:P_R_prime_fit}
shows reasonable convergence with increasing $L$).

While we have latterly focussed on the MBL regime, it was pointed out above
 that in the ergodic phase -- e.g.\   $W=1,2$  in Fig.\ \ref{fig:Average_R_t} -- the average
$\langle R\rangle (t)$ shows strong $L$-dependence at times $t\simeq 1$; reflecting incipient multifractality in the wavefunction, which is arrested at later times as the system crosses over to characteristic ergodic behaviour. Naturally, such behaviour around $t\simeq 1$ is equally apparent in the full $P_{R}(x)$ distributions for the ergodic phase shown in Fig.\ \ref{fig:P_R}.
Accordingly, the right panel in Fig.\ \ref{fig:P_R_prime_fit} shows (for $W=2$) the distributions 
$P_{\tilde{R}}(x)$ for $t=1$, in direct analogy to Fig.\ \ref{fig:P_R_prime_fit} left panel. Once again the L\'evy form appears to describe the data rather well;  now with a larger tail exponent ($\alpha \simeq 7$) than for the MBL regime (such that the variance of $P_{\tilde{R}}(x)$ is finite).


\subsubsection{$\boldsymbol{P_{\mathrm{rel}}(x)~\mathrm{distribution}}$}
\label{subsection:Prel}

\begin{figure*}
\includegraphics[width=0.9\linewidth]{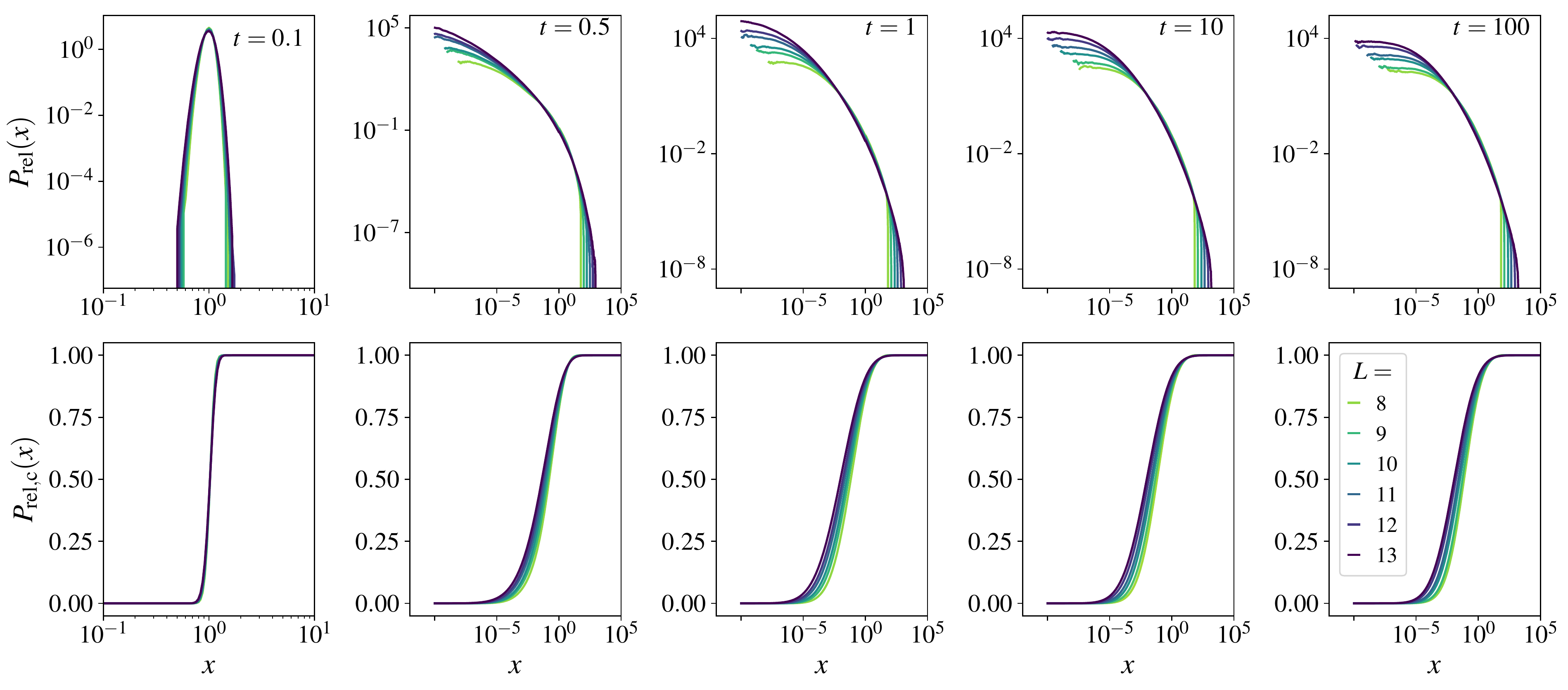}
\caption{For $W=7$. Top panels: distribution $\prel(x)$ of $x$ (Eq.\ \ref{eq:defx}) 
\emph{vs} $x$, at the sequence of $t$'s specified, and for system sizes $L$ indicated.
Bottom panels: corresponding cumulative distribution, $\prelc(x)$.
}
\label{fig:P_rel_W_7}
\end{figure*}

\begin{figure}
\includegraphics[width=\columnwidth]{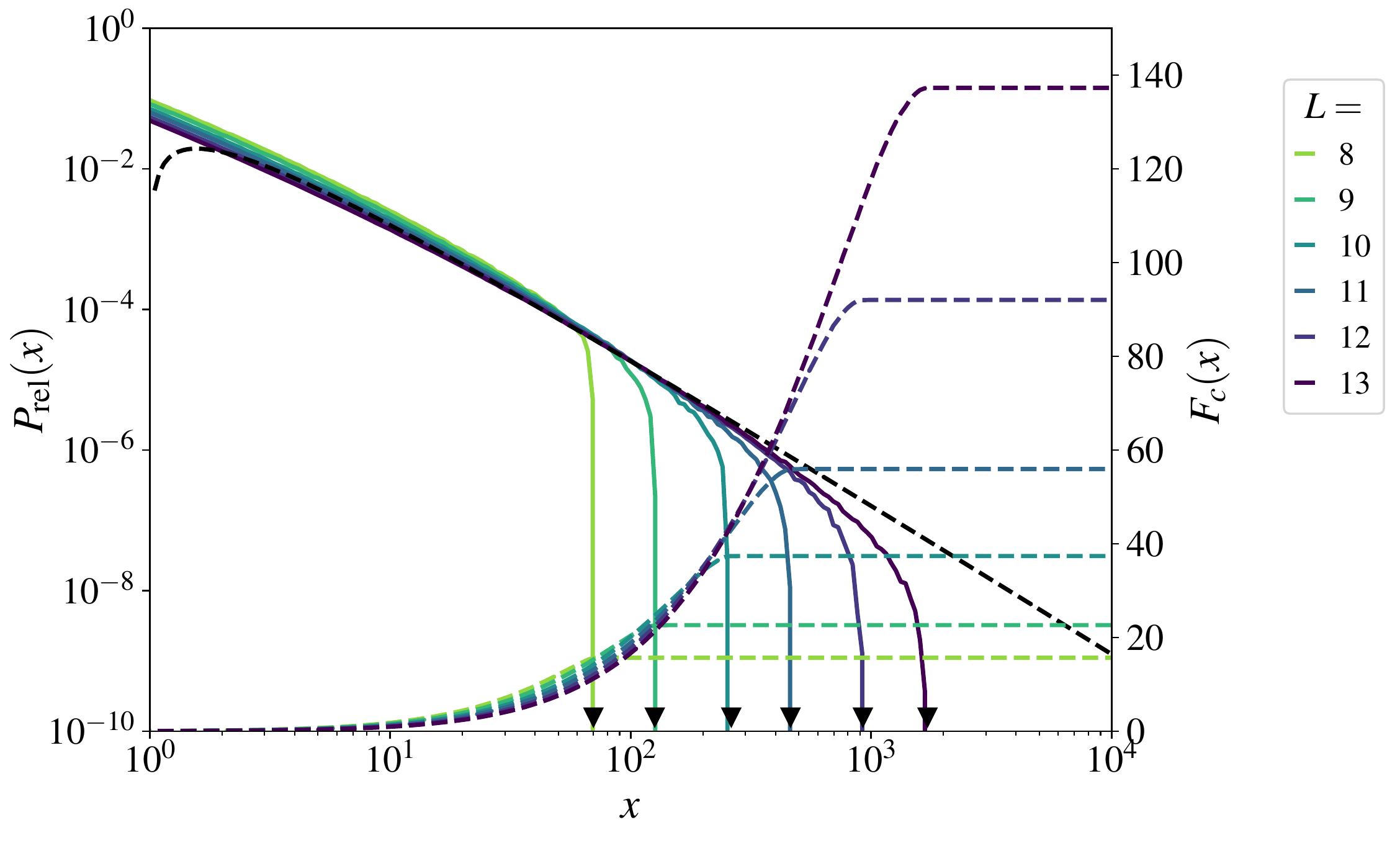}
  \caption{
	For $W=7$, with $t=100$. $F_{c}(x)$ (Eq.\ \ref{eq:Fcdef}) \emph{vs} $x$ (dashed lines, 
	right hand scale), shown for $x>1$ and different $L$ as indicated; and  $\prel(x)$ (solid lines, 
	left hand scale).	Black arrowheads show $N_{r}$ ($=\binom{L}{L/2}$ for even $L$, and 
$\binom{L}{(L\pm 1)/2}$ for odd $L$). Dashed black line shows fit to $\prel(x)$ data (see text).
	} 
	\label{fig:P_rel_MBL_fit}
\end{figure}

Complementary insight into the spread of probabilities across a row of the FS graph
comes from the distribution $\prel(x)$ (Eq.\ \ref{eq:Preldefn}). For a given row $r$,
this gives the distribution -- over disorder realisations, FS sites $J$ on the row, and initial FS sites $I$ -- of $P_{IJ}(t)$ relative to its mean value on the row, 
\begin{equation}
\label{eq:defx}
x~\equiv~\frac{P_{IJ}^{\pd}(t)}{\frac{1}{N_{r}}\underset{J:r_{IJ}=r}{\sum}P_{IJ}^{\pd}(t)}
~=~\frac{P_{IJ}^{\pd}(t)}{\frac{1}{N_{r}}P_{I}^{\pd}(r;t)}.
\end{equation}
The first moment of $\prel(x)$ is $1$ by construction, while its second moment is the
average $\langle R\rangle$ studied above.

First, consider $W=1$ (again choosing $r=L/2$). The top row of Fig.\ \ref{fig:P_rel_W_1} show
$\prel(x)$ \emph{vs} $x$ at the sequence of $t$'s specified, and for different system sizes $L$.
Corresponding cumulative distributions, $\prelc(x)=\int_{0}^{x}dy \prel(y)$, are shown in
the bottom panels. While $\prel(x)$ is not converged in $L$ for $t=1$ -- as expected from the preceding discussion -- the distributions appear converged  in $L$ for the other $t$'s shown.
The long-time  $\prel(x)$ is reached by $t=10^{2}$  (indeed essentially so by $t\sim 10$); 
and consistent with the convergence of $\prel(x)$ with  $L$, the long-time value of
$\langle R\rangle =\int dx~x^{2}\prel(x) $ is seen from  Fig.\ \ref{fig:Average_R_t} to be 
$\mathcal{O}(1)$ and $L$-independent. This long-time $\prel(x)$ is in fact rather well captured by a half-Gaussian distribution, of form $P_{G}(x)=(2/\pi)\exp(-x^{2}/\pi)$ (for $x\geq 0$), with a mean of unity and a corresponding cumulative distribution  $\mathrm{Erf}(x/\sqrt{\pi})$.
The latter is compared to the ED data in Fig.\ \ref{fig:P_rel_W_1} (final panel, dashed line),
and seen to agree well with it.

The important physical point here is that the long-time $\prel(x)$ (or $P_{G}(x)$) has a mean
of unity, and fluctuations that are also $\mathcal{O}(1)$. This means the probabilities $P_{IJ}$ are essentially uniformly distributed across the row; as evident e.g.\ from Eq.\ \ref{eq:defx} where, if all $P_{IJ}$'s on the row are comparable, then $x \sim \mathcal{O}(1)$. This 
is symptomatic of the ergodic behaviour one expects for weak disorder.

But now consider the case of $W=7$, representative of the MBL regime, for which corresponding results are shown in Fig.\ \ref{fig:P_rel_W_7}. The situation here is very different since --
particularly in the wings of $\prel(x)$ -- the distribution is clearly not converging with $L$ for any $t$ (save for $t\lesssim 0.1$, as  known on general grounds, Sec.\ \ref{section:shortt}).
This is to be expected because, as shown above, e.g.\ the long-time value of 
$\langle R\rangle =\int dx~x^{2}\prel(x)$ grows with increasing $L$, as 
$\langle R\rangle \sim N_{r}^{1-\nu}$ with exponent $\nu <1$. The question then 
is:  what features of the long-time $\prel(x)$ distribution determine that behaviour?
It must surely arise from the large-$x$ tails of $\prel(x)$
which, from Fig.\ \ref{fig:P_rel_W_7}, are `filling out' in an obvious sense with
increasing $L$.

To examine this, consider the effective cumulative distribution
\begin{equation}
\label{eq:Fcdef}
F_{c}^{\pd}(x) ~=~\int_{0}^{x}dy~y^{2}\prel(y),
\end{equation}
giving the contribution to $\langle R\rangle$ arising from different parts of the 
$\prel$ distribution. This is shown in Fig.\ \ref{fig:P_rel_MBL_fit} for $x>1$
(dashed lines and right axis), together with $\prel(x)$ itself (solid lines, left axis).
We add in passing that while the large-$x$ behaviour of $\prel(x)$ does not appear to be a pure power-law, it is quite well captured by  $\prel(x) \sim a x^{-n}\ln x$ (with $n\simeq 2.3$), shown as the black dashed line in Fig.\ \ref{fig:P_rel_MBL_fit}.
As seen from the figure, $F_{c}(x)$  for a given $L$ tends to its saturation  value at the 
$x=x_{\mathrm{m}}(L)$  for which $\prel(x)$ `crashes' in a  self-evident sense.
$x_{\mathrm{m}}(L)$ grows strongly with $L$, and the $F_{c}(x)$'s for different $L$ progressively collapse onto  an essentially common curve.

As discussed below, the maximum possible value of $x$ (Eq.\ \ref{eq:defx}) in $\prel(x)$ is in fact $N_{r}=\binom{L}{r}$ (and thus exponentially large in $L$ for any finite $r/L$).
That this is indeed the $x_{\mathrm{m}}(L)$ for which $\prel(x)$ crashes and $F_{c}(x)$ consequently plateaus is seen in Fig.\ \ref{fig:P_rel_MBL_fit}, where black arrows show
$N_{r}$. The fact that $\mathrm{max}(x)=N_{r}$ is evident from Eq.\ \ref{eq:defx}.
Since all $P_{IJ}\geq 0$ then, over the set of probabilities $P_{IJ}$ for the $N_{r}$ 
FS sites $J$ on row $r$, it arises in the case where only a single $P_{IJ}$ -- call 
it $P_{IJ^{*}}$ -- completely dominates the others (such that 
$\sum_{J:r_{IJ}=r}P_{IJ} \equiv P_{IJ^{*}}$). More generally, if the set of $P_{IJ}$ are correspondingly non-negligible for an $\mathcal{O}(1)$ number of FS sites $J$ on the row, 
then the associated $x$ is again $\mathcal{O}(N_{r})$.

As shown, it is then the large-$x$ behaviour of $\prel(x)$ which governs the
second moment $\langle R\rangle$, and consequently all higher moments, of the distribution.
Physically, this arises from FS sites  $J$ for which $P_{IJ}$  greatly exceeds the mean probability $N_{r}^{-1}\sum_{J:r_{IJ}=r}P_{IJ}$ on the row. And that of course reflects the strong inhomogeneity in the distribution of $P_{IJ}\equiv P_{IJ}(t)$ across a row, which is symptomatic of the MBL regime for sufficiently long times.

\section{Summary and discussion}
\label{sec:summary}

\begin{figure}[!t]
\includegraphics[width=\linewidth]{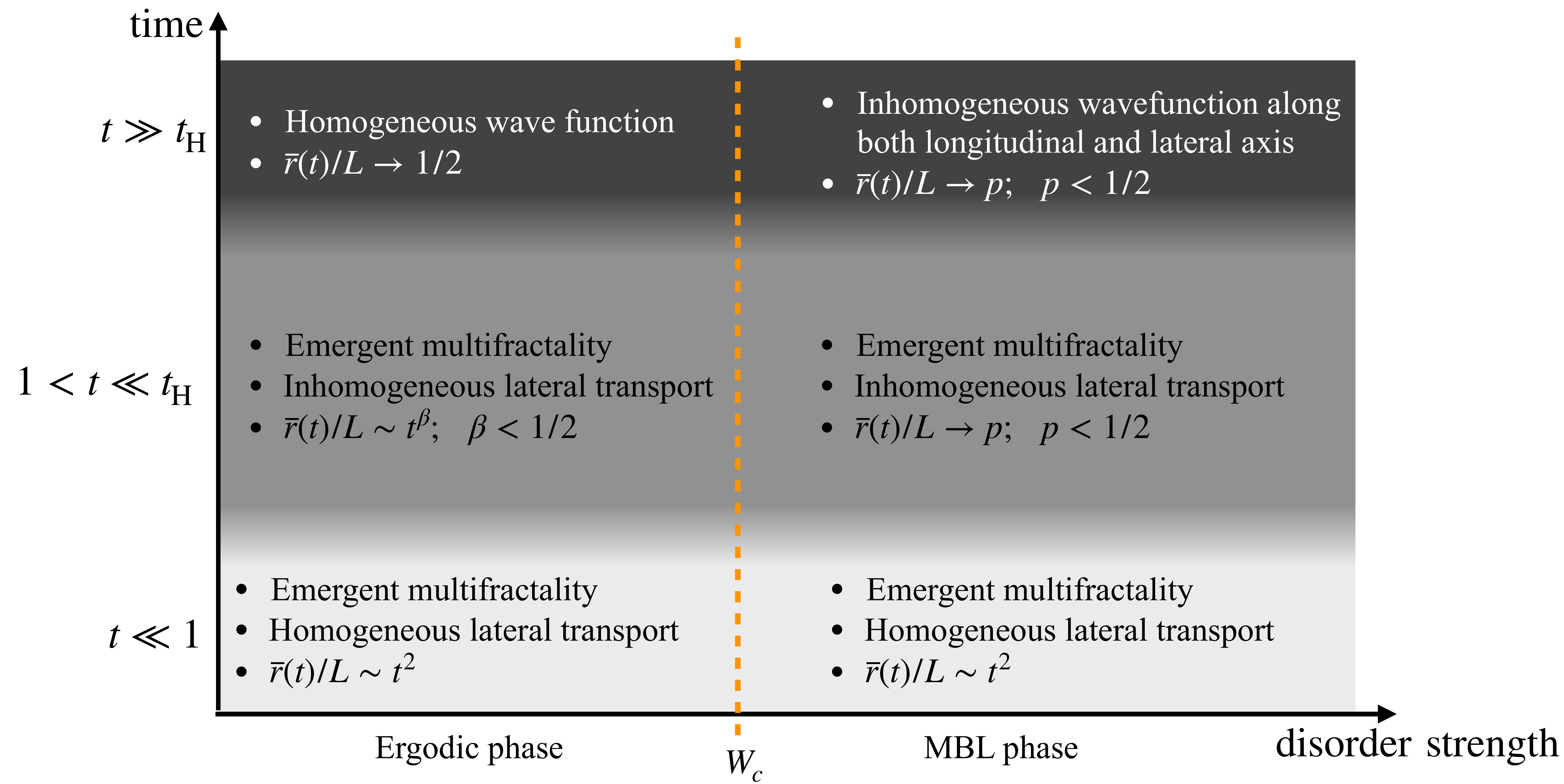}
\caption{Schematic summary of the three main time-windows in the dynamics, and their characteristic features in the ergodic and MBL phases.}
\label{fig:summary}
\end{figure}

The central question we posed at the outset was: given an initial spin configuration, how do the probability densities of the time-evolving quantum state spread out on the FS graph of a disordered quantum spin chain? In the course of investigating this question, a rather rich phenomenology was uncovered for the anatomy of probability transport on FS. This can be conveniently summarised by considering three time-windows, as follows (see Fig.~\ref{fig:summary} for a visual summary).
\begin{itemize}
    \item Short times, $t\ll 1$: In this regime, the system is agnostic to which phase it is in, ergodic or MBL. The dynamics on these scales is characterised by an emergent (multi)fractality of the time-evolving state, but homogeneous lateral probability transport across rows of the FS graph. Another crucial feature of this regime is that the emergent lengthscale $\overline{r}(t)\sim \mathcal{O}(L)$ for any finite $t$ however small; this ensures that the spin-autocorrelation $\overline{\mathcal{C}}(t)$ is necessarily less than unity. The fractal exponent $\tau_2$, as well as the lengthscale $\overline{r}(t)$, grows as $\propto t^2$ in this regime.
    \item Intermediate times, $1\lesssim t\ll t_H$: This is arguably the most interesting dynamical regime. While the emergent (multi)fractality of the entire wavefunction persists, albeit with an increasing $\tau_2$, strong inhomogeneities in the lateral probability transport also set in, reflected in (multi)fractal scalings of the row-resolved IPRs. This is also manifest in the distributions $P_R$ and $P_\mathrm{rel}$ not being converged with $L$. On these timescales, 
$\overline{r}(t)$ (or $\barrn{}(t)$) in the ergodic phase grows subdiffusively, $\sim t^\beta$ with 
$\beta<1/2$, implying an anomalous $t^{-\beta}$ power-law decay of the real-space spin autocorrelation. In the MBL phase as well, there is a power-law envelope to the decay of the spin-autocorrelation, but with clear signatures of the incipient saturation to a finite value, characteristic of that phase.

    \item Long times, $t\gtrsim t_\mathrm{H}$: This is the regime where the dynamics is essentially saturated and one sees the eigenstate properties. In the ergodic phase, the (multi)fractality gives way to a fully extended homogenenous state, both in terms of the IPR of the entire state as well as the row-resolved IPRs; and as also reflected in $\braket{R}$ saturating to an $L$-independent value, and similarly for the distributions $P_R$ and $P_\mathrm{rel}$. This is qualitatively different from the MBL regime, in which the (multi)fractality, for both the full state and at the row-resolved level, persists for arbitrarily long times. This is symptomatic of strongly inhomogeneous probability transport on the FS graph in the MBL phase,
and is also manifest in $P_R$ exhibiting a heavy-tailed L\'evy alpha-stable distribution.
\end{itemize}

While the work has presented quite a comprehensive picture of probability transport on the FS graph of a disordered quantum spin chain, it also motivates several further questions of immanent 
interest. For conserved quantities, or local observables which have a finite overlap with the former, it is worth asking if there exists a connection between potentially anomalous transport of the conserved quantity in real-space, and FS probability transport. This naturally involves space-time correlations in real space, and not just autocorrelations. Going beyond systems with conserved quantities, one can also ask about the fate of probabilty transport in the absence of \emph{any} conserved quantities, such that $\overline{r}(t)$ is not restricted to be subdiffusive nor $\overline{\mathcal{C}}(t)$ to decay as a power-law in time. 

In this work we focussed on FS probability transport, which is clearly a two-point correlation function on FS. One can generalise the question to that of the dynamics of four-point correlations on FS, with the aim of understanding entanglement growth in disordered quantum systems~\cite{bardarson2012unbounded,serbyn2013universal,lezama2019power}, both in the ergodic as well as the MBL phase.

Finally, the persistence of dynamical inhomogeneities in the MBL phase can provide us with a starting point for understanding and theorising about the role of resonances in the MBL phase from a FS point of view. Speculating that these resonances are caused by rare disorder flucuations in real space, it is also interesting to ask similar questions for MBL phases with quasiperiodic potentials, which are devoid of such rare regions~\cite{khemani2017two,khemani2017critical,roy2022diagnostics}.

\begin{acknowledgments}
This work was supported in part by EPSRC, under Grant No. EP/L015722/1 for the TMCS Centre for Doctoral Training, and Grant No. EP/S020527/1. S.R. also acknowledges support from an 
ICTS-Simons Early Career Faculty Fellowship, via a grant from the Simons Foundation
(677895, R.G.).
\end{acknowledgments}

\appendix

\section{$\boldsymbol{P_{IJ}(t)}$ Eigenstate resolution}
\label{section:eigresolution}

As mentioned in Sec.\ \ref{section:verticaltransport}, the probabilities
$P_{IJ}(t)=|\langle J|e^{-i\mathcal{H}t}|I\rangle|^{2}$ can be eigenstate resolved in the
form
\begin{equation}
\label{eq:eigres1}
P_{IJ}^{\pd}(t) ~=~\nh^{-1}\sum_{n} P_{IJ}^{(n)}(t)
\end{equation}
with the sum over eigenstates, $n$. Any quantity linear in the $\{P_{IJ}(t)\}$'s can 
thus likewise be eigenstate resolved. For example, the first moment 
$r_{I}(t)=\sum_{J}r_{IJ}P_{IJ}(t)$ is $r_{I}(t)=\nh^{-1}\sum_{n}\rn{I}(t)$
with $\rn{I}(t)=\sum_{J}r_{IJ}\PIJn{IJ}(t)$, and it is the disorder averaged 
 $\barrn{}(t)/L=\nh^{-1}\sum_{I}\barrn{I}(t)/L$ shown in Fig.\ \ref{fig:rbar(n)W=1,2};
similarly (see Eqs.\ \ref{eq:Cdef},\ref{eq:Cfull}),  
$\mathcal{C}(t)=\nh^{-1}\sum_{n}\Cn{}(t)$ with $\Cn{}(t)=1-\tfrac{2}{L}\rn{}(t)$.

Since $P_{IJ}(t)\equiv \mathfrak{Re}P_{IJ}(t)$ is pure real, Eq.\ \ref{eq:PIJamps} gives 
$P_{IJ}(t)=\sum_{n,m}\mathrm{cos}[(E_{n}-E_{m})t]A_{nI}A_{nJ}A_{mI}A_{mJ}$. Hence on 
comparison to Eq.\ \ref{eq:eigres1}, Eq.\ \ref{eq:PIJntcos} follows directly,
expressed in terms of  (real) eigenstate amplitudes ($A_{mJ}=\langle J|m\rangle$) and eigenvalues.

Considering
$P_{IJ}(t)=\mathfrak{Re}\big[\langle I|e^{i\mathcal{H}t}|J\rangle\langle J|e^{-i\mathcal{H}t}|I\rangle\big]$, and inserting the identity operator $\hat{1}=\sum_{n}|n\rangle\langle n|$,
gives
\begin{equation}
\label{eq:eigres3}
\begin{split}
\nh^{-1}P_{IJ}^{(n)}(t)=& \mathfrak{Re}\langle n|\hat{O}_{J}^{\pd}(t)\hat{O}_{I}^{\pd}|n\rangle
~~:~ \hat{O}_{J}^{\pd} =|J\rangle\langle J|
\\
=&~\tfrac{1}{2}\langle n|\big(\hat{O}_{J}^{\pd}(t)\hat{O}_{I}^{\pd}
+\hat{O}_{I}^{\pd}\hat{O}_{J}^{\pd}(t)\big)|n\rangle
\end{split}
\end{equation}
with the operator $\hat{O}_{J}=|J\rangle\langle J|$ thus defined (a so-called behemoth operator~\cite{KhaymovichBehemoth2019}); showing that $\PIJn{IJ}(t)$ can equally be expressed as an eigenstate expectation value of the self-adjoint operator, 
$\hat{O}(t)=\tfrac{1}{2}\big(\hat{O}_{J}(t)\hat{O}_{I}+\hat{O}_{I}\hat{O}_{J}(t)\big)$.

While any $P_{IJ}(t)$ itself is non-negative for \emph{all} $t$, we add that the same is not guaranteed for $\PIJn{IJ}(t)$; though it is  obvious from Eq.\ \ref{eq:PIJntcos} that this does hold in the short- and long-time limits, for which
$\nh^{-1}\PIJn{IJ}(t=0)=\delta_{IJ}A_{nI}^{2}$ and 
$\nh^{-1}\PIJn{IJ}(t=\infty)=A_{nI}^{2}A_{nJ}^{2}$. In practice this is however of little import, with quantities such as $\overline{r}^{(n)}(t)/L$ shown in Fig.\ \ref{fig:rbar(n)W=1,2} found to be non-negative for all $t$, as expected physically.

\section{Heisenberg times}
\label{section:Heistimes}

The mean level spacing at energy $\w$ is $[\nh D(\w)]^{-1}$, with $D(\w)$ the density of
states/eigenvalues normalised to unity over $\w$.  Reflecting the central limit theorem, $D(\w)$ is known to be a Gaussian,~\cite{logan2019many,welsh2018simple} with vanishing mean for the disordered TFI model under consideration, and a variance $\mu_{E}^{2} \propto L$ given 
exactly by~\cite{roy2020fock}
$\mu_{E}^{2}=L\big[J^{2}+\tfrac{1}{3}(\delta J)^{2}+\tfrac{1}{3}W^{2}+\Gamma^{2}\big]$.
The Heisenberg time  $\tH$ is the inverse of the mean level spacing.
We consider it at the band centre, $\w =0$ (where it is largest), so
$\tH =\nh D(0)=\nh/\sqrt{2\pi \mu_{E}^{2}}$ and hence
\begin{equation}
\label{eq:AppHeis1}
\tH
~=~\frac{2^{L}}{\sqrt{2\pi L\big[J^{2}+\tfrac{1}{3}(\delta J)^{2}+\tfrac{1}{3}W^{2}+\Gamma^{2}\big]}}.
\end{equation}
For all ED calculations, $J=1, \delta J=0.2$ and $\Gamma \equiv 1$ are fixed.
$\tH$ obviously increases with $L$ and decreases with disorder strength $W$. 
For $W=1,2$, and $L \in [8,14]$, $\tH$ ranges from $\sim 20$ to $\sim 10^{3}$,
while for $W=6,7$ it correspondingly ranges from $\sim 10$ to $\sim 400$.

\section{$\boldsymbol{\mblo}$}
\label{section:mblo}

We outline basic steps underlying the results given in Sec.\ \ref{subsection:mblo1}
for $\mblo$, which corresponds to the non-interacting limit $J_{\ell}=0$ of $\mathcal{H}$,
Eq.\ \ref{eq:ham}. The Hamiltonian in this case is site-separable,
$\mathcal{H}=\sum_{\ell=1}^{L}\mathcal{H}_{\ell}$, with
$\mathcal{H}_{\ell}=h_{\ell}\hat{\sigma}_{\ell}^{z}+\Gamma \hat{\sigma}_{\ell}^{x}$.
The latter is diagonalised as
\begin{equation}
\label{eq:mbloapp1}
\mathcal{H}_{\ell}^{\pd}~=~\phi_{\ell}^{\pd}\hat{\tilde{\sigma}}_{\ell}^{z}
~~~~:~\phi_{\ell}^{\pd}=\sqrt{h_{\ell}^{2}+\Gamma^{2}}
\end{equation}
in terms of the spin-$1/2$ operator
\begin{equation}
\label{eq:mbloapp2}
\hat{\tilde{\sigma}}_{\ell}^{z}~=~\frac{h_{\ell}\hat{\sigma}_{\ell}^{z}+\Gamma \hat{\sigma}_{\ell}^{x}}{\sqrt{h_{\ell}^{2}+\Gamma^{2}}}
\end{equation}
(such that $[\hat{\tilde{\sigma}}_{\ell}^{z}]^{2}=1$).
An eigenstate $|n\rangle$ of $\mathcal{H}$ is simply a product state
of the set of $\tilde{\sigma}$-spins, $|n\rangle =|\{\tilde{\sigma}_{\ell}^{z}\}\rangle$ with
each $\tilde{\sigma}_{\ell}^{z}$ either $+1$ or $-1$.

Now consider the probability amplitude $G_{IJ}(t)=\langle J| e^{-i\mathcal{H}t}|I\rangle$
(with a general FS site $|K\rangle \equiv |\{S_{\ell,K}\}\rangle$ in the notation specified
in Sec.\ \ref{section:model}). Since $\mathcal{H}$ is site-separable, $G_{IJ}(t)$ is a separable product,
\begin{equation}
\label{eq:mbloapp3}
G_{IJ}^{\pd}(t)  ~=~\langle J| e^{-i\mathcal{H}t}|I\rangle
~=~\prod_{\ell =1}^{L}\langle S_{\ell,J}^{\pd}| e^{-i\mathcal{H}_{\ell}^{\pd}t}|S_{\ell,I}^{\pd}\rangle ,
\end{equation}
and
$e^{-i\mathcal{H}_{\ell}t}=\mathrm{cos}(\phi_{\ell}t) -i\hat{\tilde{\sigma}}_{\ell}^{z}\mathrm{sin}(\phi_{\ell}t)$. The matrix elements in the product are readily evaluated, 
\begin{equation}
\label{eq:mbloapp4}
\begin{split}
&\langle S_{\ell,J}^{\pd}| e^{-i\mathcal{H}_{\ell}^{\pd}t}|S_{\ell,I}^{\pd}\rangle
~=~
\\
&
\begin{cases}
\mathrm{cos}(\phi_{\ell}^{\pd}t) -\frac{ih_{\ell}}{\sqrt{h_{\ell}^{2}+\Gamma^{2}}}S_{\ell,I}^{\pd}
\mathrm{sin}(\phi_{\ell}^{\pd}t)
~:~S_{\ell,J}^{\pd}=S_{\ell,I}^{\pd}\\
-\frac{i\Gamma}{\sqrt{h_{\ell}^{2}+\Gamma^{2}}}\mathrm{sin}(\phi_{\ell}^{\pd}t)
~~~~~~~~~~~~~~~:~S_{\ell,J}^{\pd}=-S_{\ell,I}^{\pd}
\end{cases}
\end{split}
\end{equation}
according to whether the local spin $S_{\ell,J}=\pm S_{\ell,I}$.

Let the FS sites $J,I$ be separated by a Hamming distance $r_{IJ}=r$. 
Then by definition $r$ real-space sites  have $S_{\ell,J}=-S_{\ell,I}$, while $(L-r)$ sites 
have $S_{\ell,J}=+S_{\ell,I}$. Eqs.\ \ref{eq:mbloapp3},\ref{eq:mbloapp4} then give
\begin{equation}
\nonumber
\begin{split}
G_{IJ}^{\pd}(t)  =&\prod_{\ell \in r}\Big[\frac{-i \Gamma}{\sqrt{h_{\ell}^{2}+\Gamma^{2}}}\mathrm{sin}(\phi_{\ell}^{\pd}t)\Big]
\\
&\times
\prod_{\ell \in (L-r)}\Big[\mathrm{cos}(\phi_{\ell}^{\pd}t) -\frac{i h_{\ell}}{\sqrt{h_{\ell}^{2}+\Gamma^{2}}}S_{\ell,I}^{\pd}
\mathrm{sin}(\phi_{\ell}^{\pd}t)\Big]
\end{split}
\end{equation}
in an obvious notation. From this (recalling $[S_{\ell,I}]^{2}=1$) $P_{IJ}(t)=|G_{IJ}(t)|^{2}$ follows,
\begin{equation}
\label{eq:mbloapp5}
\begin{split}
P_{IJ}^{\pd}(t)=&\prod_{\ell \in r}\Big[\frac{\Gamma^{2}}{h_{\ell}^{2}+\Gamma^{2}}\mathrm{sin}^{2}(\phi_{\ell}^{\pd}t)\Big]
\\
&\times
\prod_{\ell \in (L-r)}\Big[ 1-\frac{\Gamma^{2}}{h_{\ell}^{2}+\Gamma^{2}}\mathrm{sin}^{2}(\phi_{\ell}^{\pd}t)\Big].
\end{split}
\end{equation}
This can now be averaged over disorder realisations, and since the random fields
$\{h_{\ell}\}$ are i.i.d.,  
\begin{equation}
\label{eq:mbloapp6}
\overline{P}_{IJ}^{\pd}(t)=\Big\langle\frac{\Gamma^{2}}{h_{\ell}^{2}+\Gamma^{2}}\mathrm{sin}^{2}(\phi_{\ell}^{\pd}t)
\Big\rangle_{\dis}^{r}\times
\Big\langle 1-\frac{\Gamma^{2}}{h_{\ell}^{2}+\Gamma^{2}}\mathrm{sin}^{2}(\phi_{\ell}^{\pd}t)\Big\rangle_{\dis}^{(L-r)}
\end{equation}
which is Eqs.\ \ref{eq:PIJmblo},\ref{eq:z0def} as required.
Eq.\ \ref{eq:mbloapp6} is indeed seen to depend solely on the Hamming distance $r_{IJ}=r$
between FS sites $J,I$; such that, from Eq.\ \ref{eq:Prtdef},
$\overline{P}(r;t)\equiv \overline{P}_{I}(r;t) \equiv \binom{L}{r}\overline{P}_{IJ}(t)$,
as given explicitly in Eq.\ \ref{eq:Pbarmblo}.

Eq.\ \ref{eq:Pbarmblo} can obviously be cast in the form
\begin{equation}
\label{eq:mbloapp7}
\overline{P}(r;t)~=~\binom{L}{r}
\big[1+e^{-1/\xi_{F}^{0}(t)}\big]^{-L} e^{-r/\xi_{F}^{0}(t)}
\end{equation}
in terms of a correlation length $\xi_{F}^{0}(t)$ defined by
$1/\xi_{F}^{0}(t) = \ln(\tfrac{1}{z_{0}(t)}-1)$. This is the $\mblo$ counterpart of the 
short-time result Eq.\ \ref{eq:PIJrshorttXi} (in the latter case, 
$\overline{P}(r;t)\equiv \binom{L}{r}P_{IJ_{r}}(t)$).
Eq.\ \ref{eq:PIJrshorttXi} itself is of course general -- in the sense that it holds for all interaction and disorder strengths -- and Eq.\ \ref{eq:mbloapp7} correctly reduces to it for 
$\Gamma t \ll 1$. 

We also point out the connection between the long-time limit 
$\overline{P}(r;t=\infty)$ of Eq.\ \ref{eq:mbloapp7}, and the eigenstate correlation function 
$\overline{F}_{n}(r)$ defined generally by Eq.\ \ref{eq:Fnbardef} and given in terms of FS correlation lengths $\xi_{F,n}$ for eigenstates $n$ by Eq.\ \ref{eq:Fnbarinf}.
$\overline{P}(r;\infty)$ is given generally by
$\overline{P}(r;\infty)=\nh^{-1}\sum_{n}\overline{F}_{n}(r)$.
For $\mblo$ one can however show that the disorder-averaged  $\overline{A_{nI}^{2}A_{nJ}^{2}}$ 
is independent of the particular eigenstate $n$. From Eq.\ \ref{eq:Fnbardef}, $\overline{F}_{n}(r)$ is thus independent of $n$, whence $\overline{P}(r;\infty)\equiv \overline{F}_{n}(r)$ 
gives the connection sought.

Comparison of Eq.\ \ref{eq:mbloapp7} for $t=\infty$ to Eq.\ \ref{eq:Fnbarinf},
$\overline{F}_{n}(r)=\binom{L}{r}(1+e^{-1/\xi_{F,n}})^{-L}e^{-r/\xi_{F,n}}$,
then relates directly the infinite-$t$ dynamical correlation length $\xi_{F}^{0}(\infty)$ to the
($n$-independent) eigenstate correlation length, viz.\ $\xi_{F}^{0}(t=\infty)\equiv \xi_{F,n}$;
given explicitly (using Eq.\ \ref{eq:z0itop}) by
$\xi_{F}^{0}(\infty)=[\ln(\tfrac{1}{p}-1)]^{-1}$, with 
$\xi_{F}^{0}(\infty)\propto 1/\ln W$ for $W\gg 1$.



\bibliography{refs}

\end{document}